 \documentclass[ reprint, twocolumn,  amsmath,
 amssymb, aps, amssymb]{revtex4-2}
\usepackage{amssymb}
\usepackage{graphicx}
\usepackage{dcolumn}
\usepackage{bm}%
\usepackage{gensymb}

\usepackage[dvipsnames]{xcolor}
\usepackage{ragged2e}

\renewcommand{\thefigure}{\textbf{\arabic{figure}}}
\usepackage{amssymb}
\usepackage{amsmath}

\usepackage{color}
\usepackage{gensymb}
\usepackage{microtype}
\usepackage{dirtytalk}
\definecolor{red}{rgb}{1,0,0}
\definecolor{blue}{rgb}{0,0,1}
\definecolor{black}{rgb}{0,0,0}
\definecolor{green}{rgb}{0.13, 0.55, 0.13}
\definecolor{goldenrod}{rgb}{0.85, 0.65, 0.13}
\definecolor{green2}{rgb}{0.0, 1.0, 0.0}
\definecolor{brown(web)}{rgb}{0.65, 0.16, 0.16}
\definecolor{purple}{rgb}{0.8, 0.6, 0.8}
\usepackage{xr-hyper}
\usepackage{hyperref}
\usepackage{ulem}
\usepackage{soul}

\begin{document}

\title{Tuning the bulk behavior and 2D interfacial self-assembly of microgels by Keggin-type polyoxometalate ionic specificity}
\author{Antonio Rubio-Andrés}
\author{Delfi Bastos-González}
\author{Miguel Angel Fernandez-Rodriguez\textsuperscript{*}}

\affiliation{Laboratory of Surface and Interface Physics, Biocolloid and Fluid Physics Group, Department of Applied Physics, Faculty of Sciences, University of Granada, Granada, 18071, Spain\\
E-mail: mafernandez@ugr.es}


\begin{abstract}

Finding new ways to tune the behavior of thermoresponsive microgels in bulk and confined at 2D liquid interfaces is key to achieve a deeper understanding and control of these smart materials. We studied the interaction of positively charged pNIPAM microgels with the Keggin-type polyoxometalate $Na_{3}PW_{12}O_{40}$ (POM). In bulk, we observed charge inversions below and above the volume phase transition temperature (VPTT) at significantly low POM concentrations as $5\cdot10^{-5}$ M. In the presence of POM, the microgels exhibited a deswelling-swelling-deswelling behaviour below the VPTT, and a two-step further deswelling above the VPTT. When microgels were confined at 2D water/air interfaces, adding $10^{-5}$ M of POM below the VPTT was equivalent to heat above the VPTT and compress the monolayer from $5$ to $20\,\text{mN m$^{-1}$}$. Above the VPTT, the diameter at the interface did not change while the portion immersed in the subphase further deswelled, in agreement with the behavior in bulk. Adding more POM did not change the diameter at the interface nor the height of the microgels, showing a saturation effect in 2D. The restructuring of the pNIPAM polymeric network by the POM was characterized by EDS mapping and XPS. The microgel monolayers with POM improved their resistance to plasma etching, which could be useful for soft colloidal lithography.

\end{abstract}






\maketitle

\section*{Introduction}
\label{introduction}

Microgels are soft colloidal micro- and nanoparticles composed of cross-linked polymers that are swollen in a good solvent. Poly-(N-isopropylacrylamide) (pNIPAM) microgels dispersed in water are widely used thanks to their thermoresponsiveness, exhibiting a Volume Phase Transition Temperature (VPTT). The microgel swells below the VPTT and deswells above it, expelling water and stiffening in the process. \cite{lyon2012polymer}. It is possible to tune their response to other external stimuli during the synthesis, e.g. they can become pH-responsive by adding amphoteric co-monomers \cite{naseem2019adsorptive, schmidt2020influence}. 

One of the usual synthesis routes to obtain pNIPAM microgels is the precipitation polymerization, where they develop a Gaussian profile on the cross-linking density, with a highly cross-linked core and a less cross-linked corona \cite{geisel2014compressibility}. When a microgel adsorbs at a liquid interface, the portion in contact with the interface stretches due to the surface tension, only counterbalanced by the internal elasticity of the polymeric network, which it is proportional to the cross-linking density. As a result, the microgel exhibits a \textit{fried-egg} shape when it is seen from above the interface, with the portion below the subphase still well solvated \cite{camerin2019microgels}.
    
When confined at interfaces, microgels can be used as Pickering emulsion stabilizers, where their responsiveness allows to destabilize the emulsions by the influence of an external stimulus \cite{Fernandez_review, tatry2021pickering}. Moreover, when transferred to a silicon substrate, the deposited microgels can perform as lithography masks for soft colloidal lithography, to fabricate arrays of vertically aligned silicon nanowires \cite{fernandez2021near}. 

In order to improve their performance in these or any applications, we need to gain a deeper understanding of the behavior of microgels confined at fluid interfaces. Indeed, the portion of the microgel still immersed in the water subphase will keep its responsiveness to external stimuli, while the portion stretched at the interface will be mainly dominated by the surface tension \cite{schmidt2020influence,bochenek2019effect, bochenek2022situ}. Concerning the thermoresponsiveness, the diameter of the microgel at the interface remains the same regardless of the temperature. Nevertheless, as it occurs for microgels in bulk, the portion immersed in the water subphase deswells above the VPTT, decreasing the diameter of that portion \cite{bochenek2019effect, bochenek2022situ,harrer2019stimuli, vialetto2022novel}. Upon deposition on a substrate, this deswelling results in an increase of the microgel height \cite{bochenek2019effect}.

Recently, Schmidt et al. showed that for pH-responsive microgels confined at fluid interfaces the behavior is different upon pH-swelling/deswelling depending on the size of the microgel \cite{schmidt2020influence}. For big microgels (e.g. 800 nm) they found a similar behavior compared to that found for the thermoresponsiveness, with invariable diameter at the interface while the portion immersed in the water subphase was still pH-responsive. Interestingly, for small microgels (e.g. 250 nm) the diameter at the interface decreased upon pH-swelling of the microgels.
    
Another tool that can be used to tune the properties of microgels in dispersion, e.g. their charge and size, is the addition of salt. While this provides electrostatic screening of charges, we will particularize here to ionic specific interactions. The influence of the ionic specificity on pNIPAM has been widely analyzed in bulk for both free chains and microgels \cite{zhang2005specific, lopez2014thermally, daly2000study}. In both cases, the properties of pNIPAM are deeply affected by salts belonging to the Hofmeister series, specially when anions act as counter-ions \cite{perez2015anions, bastos2016ions}. However, there is a lack of experimental works studying how salts influence the interfacial properties of pNIPAM, probably due to the difficulty in obtaining reliable and accurate data in 2D systems.

There are many interesting ions candidates with complex behaviors, such as cobaltabisdicarbollide anions, but these exhibit interfacial activity \cite{COSAN}, and therefore they would compete with the adsorption of microgels at interfaces. An interesting kind of anions are the Keggin-type polyoxometalates ($\text{POMs}$), nanosized metal-oxide clusters with high valences and well defined structures \cite{kruse2022polyoxometalate}. They do not exhibit interfacial activity by themselves, but they can adsorb at the interface in the presence of interfacially active molecules \cite{ hohenschutz2020nano, girard2019thermodynamic}. Moreover, they can induce the self-assembly of free pNIPAM chains in bulk into sheets and globules \cite{kruse2022polyoxometalate, buchecker2019self}, but a comprehensive study on how this effect translates to microgels, both in bulk and at interfaces, is only recently starting to be studied \cite{simons2023tailoring}. 

This strong interaction of POMs with pNIPAM has been observed also with proteins and seems difficult to be modeled by classical models. Recently, it has been proposed a quantum description of the interaction \cite{POM_QM}. The parallelism with proteins comes from the amide groups of pNIPAM that resemble the aminoacid structure. Furthermore, $\text{POMs}$ have been gaining attention due to their interesting and versatile applications, from their use in catalytic reactions \cite{liu2021polyoxometalate}, anticancer drugs \cite{bijelic2019polyoxometalates}, hybrid antimicrobial materials \cite{soria2021pom}, or pollution removal \cite{xu2020green} to cite some applications. 

In this work, we study the effect of the Keggin-type $\text{POM}^{3-}$ anion $[PW_{12}O_{40}]^{3-}$ on the behavior of pNIPAM microgels in bulk, by means of laser Doppler microelectrophoresis and Dynamic Light Scattering (DLS) size measurements, and at the water/air interface, from monolayer deposition and AFM, TEM, STEM-HAADF, EDS and XPS characterizations. Finally, we briefly explore a potential application in Soft Colloidal Lithography of the pNIPAM microgel monolayers deposited in the presence of the $\text{POM}^{3-}$ by testing their resistance to plasma etching. 

\section*{Materials and Methods}
\label{Materials_methods}

\subsection*{Materials}

The pNIPAM microgels were synthesized as reported in a previous work by precipitation polymerization \cite{perez2015anions}. N-Isopropylacrylamide (NIPAM), aminoethyl methacrylate hydrochloride (AEMH), N-methylenebisacrylamide (BIS) and $\text{2,{2}'-azobis}$ (2-amidinopropane) dihydrochloride (V50) were obtained from Acros in analytical grade and used as received. The proportions relative to NIPAM were: AEMH at 2.6 wt\%, BIS at 1.6 wt\%, and V50 at 1.1 wt\%. The positive charge of the microgel comes from the amino groups in the AEMH and V50. The $\text{POM}^{3-}$ $Na_{3}PW_{12}O_{40}$ (Sigma Aldrich, 99.9$\%$) and NaCl (Sigma Aldrich, 99$\%$) were used as received. The different solutions, ranging from $10^{-7}$ M to $10^{-1}$ M were prepared by adjusting the solutions to pH 3 with $HCl$ (Sigma Aldrich, 37$\%$) since at this pH the polyoxometalate ($\text{POM}$) was stable \cite{ammam2013polyoxometalates}. Isopropyl alcohol (Sigma Aldrich, 99.8$\%$) was used as extension agent for the experiments at the water/air interfaces. Silicon substrates of 2 x 1 $\text{cm}^{2}$ with $\left \langle 100 \right \rangle$ orientation (p-type, Boron doped, 1-10 $\Omega \: \text{cm}$ from University Wafer Inc., USA) were cut by laser (Laser E-20 SHG II, Rofin, USA) and used for the deposition of microgel monolayers without further modification.

\subsection*{Physicochemical Characterization in bulk}

The hydrodynamic diameter ($D_{h}$) and the electrophoretic mobility ($\mu_{e}$) were measured via a ZetaSizer NanoZ (Malvern Instruments, UK). All the measurements were performed at temperatures between 25 \degree C and 50 \degree C to investigate the swollen and deswollen states of the microgels, respectively. The results were obtained by the average of three measurements and their corresponding standard deviations. For each characterization, a microgel solution of 0.1 wt$\%$ was prepared 15 minutes prior to the measurement. When the desired temperature was achieved, the sample was stabilized during 3 minutes before starting the experiments. The VPTT was obtained from a sigmoidal fit of the evolution of the $D_{h}$ as a function of the temperature in the absence of salt, as detailed in the SI (see Figure \ref{fig:SupportingCaracterizacion}).

\subsection*{Langmuir trough depositions}

Details on silicon substrate sample preparation, cleaning protocols, Langmuir trough (KSV NIMA, Biolin Scientific, Sweden) barrier compression and dipping speeds are detailed in the SI. First, a silicon substrate was attached to the dipper arm of the Langmuir trough and lowered below the interface position at an angle of 60\degree. The clean Langmuir trough was then filled with the desired salt solution at pH 3, and heated above the VPTT when needed thanks to a thermal bath heating the trough from outside. When the interface was considered clean enough (see SI), the microgel suspension was deposited at the interface thanks to a 100 $\mu L$ glass microsyringe, up to a surface pressure of $\Pi \simeq 1\,\text{mN m$^{-1}$}$, allowing it to equilibrate for 10 minutes. Next, we performed two types of experiments. The first type was intended to characterize the compression curve of the microgel. In this case the monolayer was compressed while simultaneously lifting the substrate across the interface. Just before the substrate completely crossed the interface, the barriers were fully opened to create a sudden change in $\Pi$ which we used as a reference point to match each position on the substrate with its corresponding surface pressure via AFM \cite{bochenek2019effect}. In this way, it is possible to reconstruct the compression curve by representing $\Pi$ as a function of the area occupied by each microgel. The other type of experiment was intended to investigate the effect of the $\text{POM$^{3-}$}$ concentration on the self-assembly of the microgel monolayer. First, the monolayer was compressed up to $\Pi = 5\,\text{mN m$^{-1}$}$ and deposited on half of the substrate. Next, the monolayer was compressed up to $\Pi = 20\,\text{mN m$^{-1}$}$ and deposited on the rest of the substrate. In this way, there were two different areas over the substrate corresponding to the two values of $\Pi$. After each deposition experiment, the Langmuir trough was cleaned as detailed in SI.

\subsection*{AFM and image analysis}

The deposited monolayers were characterized by means of Atomic Force Microscopy (AFM, motorized Dimension 3000) in tapping mode (Tap300Al-G cantilevers, 300 kHz, 40 N/m, BudgetSensors, Bulgaria). We acquired 512 x 512 pixels$^{2}$ images of 40 x 40 $\mu$m$^{2}$ and 10 x 10 $\mu$m$^{2}$ over the microgel monolayers deposited on the silicon substrates. For the compression curve, we acquired images every 500 $\mu $m over the substrate from the reference point thanks to a customized motorization of the AFM \cite{hardwareX}. For the substrates with two regions at 5 and 20 $\text{mN m$^{-1}$}$, we acquired images of each region. The images were post-processed with the software Gwyddion to level them and increase contrast. Next, the images were converted to 8-bits grey-scale images and analyzed using a customized particle tracking software based on the Python version of the publicly available particle tracking code by Crocker and Grier TrackPy \cite{crocker1996methods}. Thus, we localized the center of each particle excluding the microgels placed in the edges of the image, and calculated their radial distribution functions g(r), and the nearest neighbor distance.

\subsection*{TEM characterization and chemical identification}

In order to image the microgel monolayer and identify  the chemical species present in it, we performed TEM, High-Angle Annular Dark-Field imaging (STEM-HAADF), and Energy-Dispersive X-ray Spectroscopy (EDS) measurements (Thermo Fisher Scientific TALOS F200X) with an acceleration voltage of 200 kV. To reproduce the deposition procedure described in the previous section, microgels were deposited from the water/air interface at $\Pi \simeq 20$ $\text{mN m$^{-1}$}$ into a copper TEM grid, and dried at room temperature. While we cannot ensure that $\Pi$ is exactly the same as for silicon substrates due to the deposition process, this serves as a first approximation to characterize self-assembled monolayers. The details on the sample preparation are presented in Figure \ref{fig:SupportingPreparationTEM}.

\subsection*{XPS measurements}

The surface chemistry of the deposited monolayers was characterized by means of X-ray Photoelectron Spectroscopy (XPS, PHI VersaProbe II equiped with a monochromated Al X-Ray source for excitation at 1486.6 eV) with an analyzing spot diameter of 200 $\mu$m. The data treatment and peak fitting was done using the \textit{XPSPeak} software.

\subsection*{Plasma etching}

Air plasma etching was applied to the deposited microgels to test their resistance to plasma etching in the presence of $\text{POM}^{3-}$. Measurements were done by means of a radio frequency plasma device (KX1050 Plasma Asher, Emitech). Substrates with deposited microgels were subjected to air plasma during 30 minutes at 100W and characterized by AFM before and after the treatment.
   
\section*{Results and Discussion}
\label{results_discussion}
\begin{figure}
\begin{center}
\includegraphics[width=\linewidth]{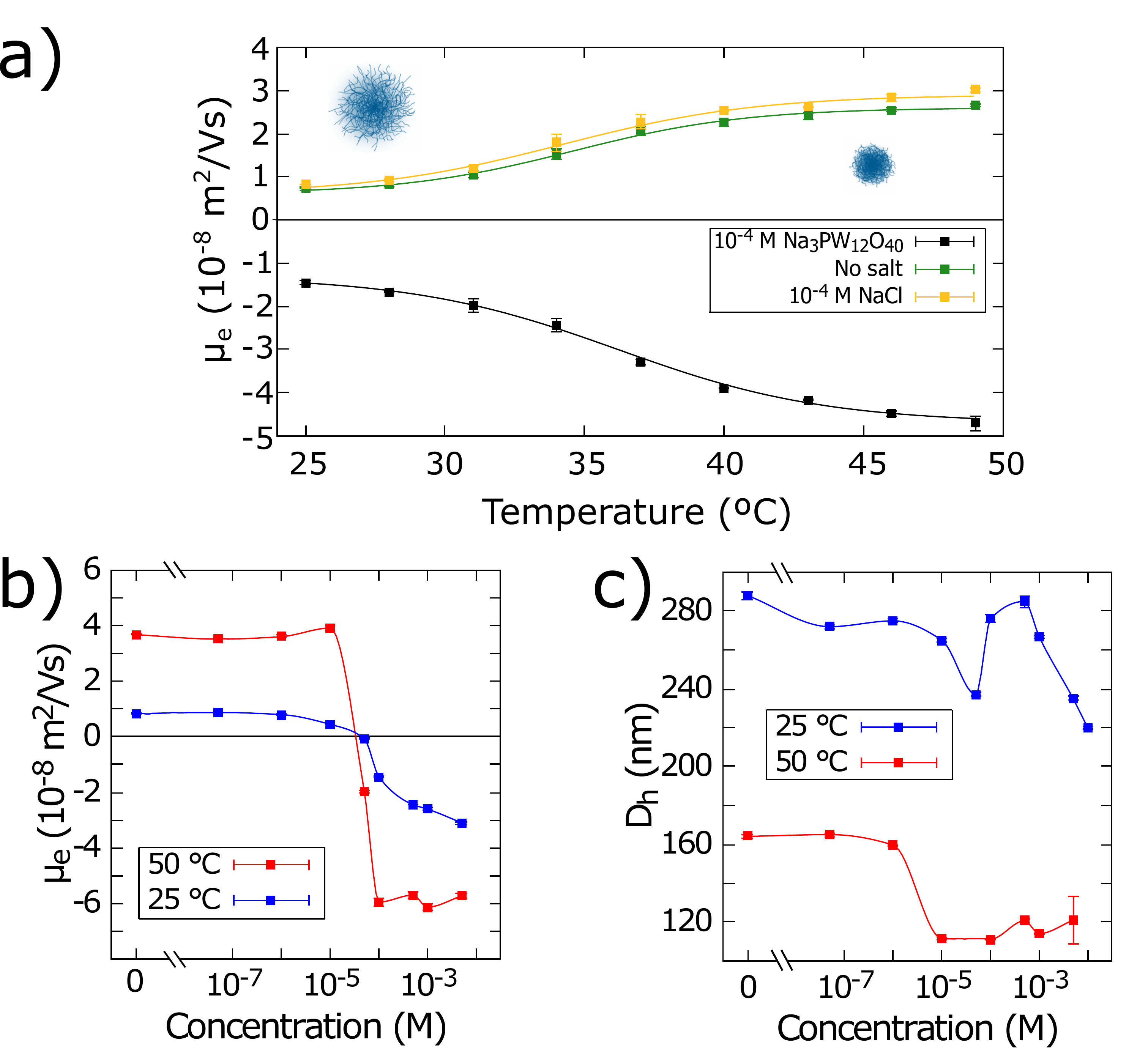}
\caption{\textbf{a)} Electrophoretic mobility ($\mu_{e}$) as a function of temperature for positively charged pNIPAM microgels at pH 3 in the absence of salt (\textcolor{green}{$\blacksquare$}), with $10^{-4}$ M of NaCl (\textcolor{goldenrod}{$\blacksquare$}), and $10^{-4}$ M of $\text{POM}^{3-}$, ({$\blacksquare$}). \textbf{b-c)} $\mu_{e}$ and hydrodynamic diameter $D_{h}$ for the same microgels as in a) as a function of the concentration of $\text{POM}^{3-}$  at 25 \degree C (\textcolor{blue}{$\blacksquare$}), and at 50 \degree C (\textcolor{red}{$\blacksquare$}). The lines are guides to the eye.}
\label{fig:Figura1}
\end{center}
\end{figure}

Characterizing the electrophoretic mobility, $\mu_{e}$, of a microgel dispersion is useful to determine the accumulation of different type of ions on the microgels. This allows to compare how the electric state of the surface of the microgels changes depending on the nature of the ions present in solution \cite{bastos2016ions}. In Figure \ref{fig:Figura1}a we present the $\mu_{e}$ trend from 25 \degree C to 49 \degree C for a microgel dispersion in the presence of $10^{-4}$ M of $\text{POM}^{3-}$. In order to make comparisons, we also performed this characterization in the absence of salt, and with $10^{-4}$ M of NaCl. The green curve without salt at pH 3 shows positive $\mu_{e}$ values as expected for the positive charge of our microgels, with no significant changes after adding $10^{-4}$ M of NaCl. Around the VPTT, $33.7 \pm 0.2$ \degree C, there was an increase in $\mu_{e}$ due to the deswelling of the microgel, which resulted in more charges per unit of area. This scenario significantly changes in the presence of $\text{POM}^{3-}$ at $10^{-4}$ M. In this case, the electrophoretic mobility inverted its sign, which means a charge inversion of the pNIPAM microgels in all range of temperatures studied. The negative electrokinetic charge in the presence of $\text{POM}^{3-}$ increases as the microgel deswells, reaching a value even higher, in absolute value, than the positive $\mu_{e}$ reached with NaCl at $10^{-4}$ M. We previously observed charge inversion with the monovalent tetraphenil borate anion ($\text{Ph}_{4}\text{B}^{-}$) at $10^{-3}$ M, but only above the VPTT, where the microgels show a more hydrophobic nature \cite{perez2015anions}. The charge reversal in the swollen state of the pNIPAM, when the microgel is more hydrated, shows that the $\text{POM}^{3-}$ also strongly accumulates in the pNIPAM interface in its more hydrophilic state. Our results confirm that the $\text{POM}^{3-}$ anions show a strong interaction with positive pNIPAM microgels in both the more hydrophilic and more hydrophobic states, below and above the VPTT, respectively. This high ionic specificity is beyond the one of a pure electrostatic interaction and it has been ascribed to its dual character of ion and nanocolloid \cite{malinenko2018keggin, drummond2019can} or to superchaotropicity \cite{assaf2018chaotropic}. 

Next, we conducted $\mu_{e}$ and hydrodynamic diameter $D_{h}$ measurements as a function of the $\text{POM}^{3-}$ concentration. At 25 \degree C, the results in Figure \ref{fig:Figura1}b show no significant changes on the microgel charge below $10^{-5}$ M. At $5\cdot 10^{-5}$ M we found the isoelectric point indicating a full screening of the microgel charge. By increasing the $\text{POM}^{3-}$ concentration, a strong adsorption of $\text{POM}^{3-}$ was observed as the $\mu_{e}$ became negative. Above $5\cdot 10^{-5}$ M, despite existing an electrostatic repulsion between the already negative microgel and the $\text{POM}^{3-}$ anions, the value of $\mu_{e}$ kept decreasing, which we hypothesize it might be mediated by the softness of the swelled microgels that would allow for more anions to keep adsorbing onto them. As we will discuss, the $D_h$ measurements in \ref{fig:Figura1}c support this hypothesis. At 50 \degree C, the microgels deswells with the corresponding increase in charge per unit area, exhibiting the same trend in $\mu_{e}$ as at 25 \degree C. However, at $5\cdot10^{-5}$ M we already observed a charge inversion which significantly increased as the $\text{POM}^{3-}$ concentration increased up to $10^{-4}$ M. Beyond that point, further increases in $\text{POM}^{3-}$ concentration were not reflect on the $\mu_e$, which remained constant. These results seem to indicate that the interaction of the $\text{POM}^{3-}$ with the pNIPAM was even stronger in its deswollen state. These results are in good agreement with our previous studies on the adsorption of $\text{POM}^{3-}$ and charge inversion for hard nanoparticles, where we observed a stronger interaction of POMs when surfaces were more hydrophobic \cite{drummond2019can}. 

\begin{figure*}[t!]
\begin{center}
\includegraphics[width=\linewidth]{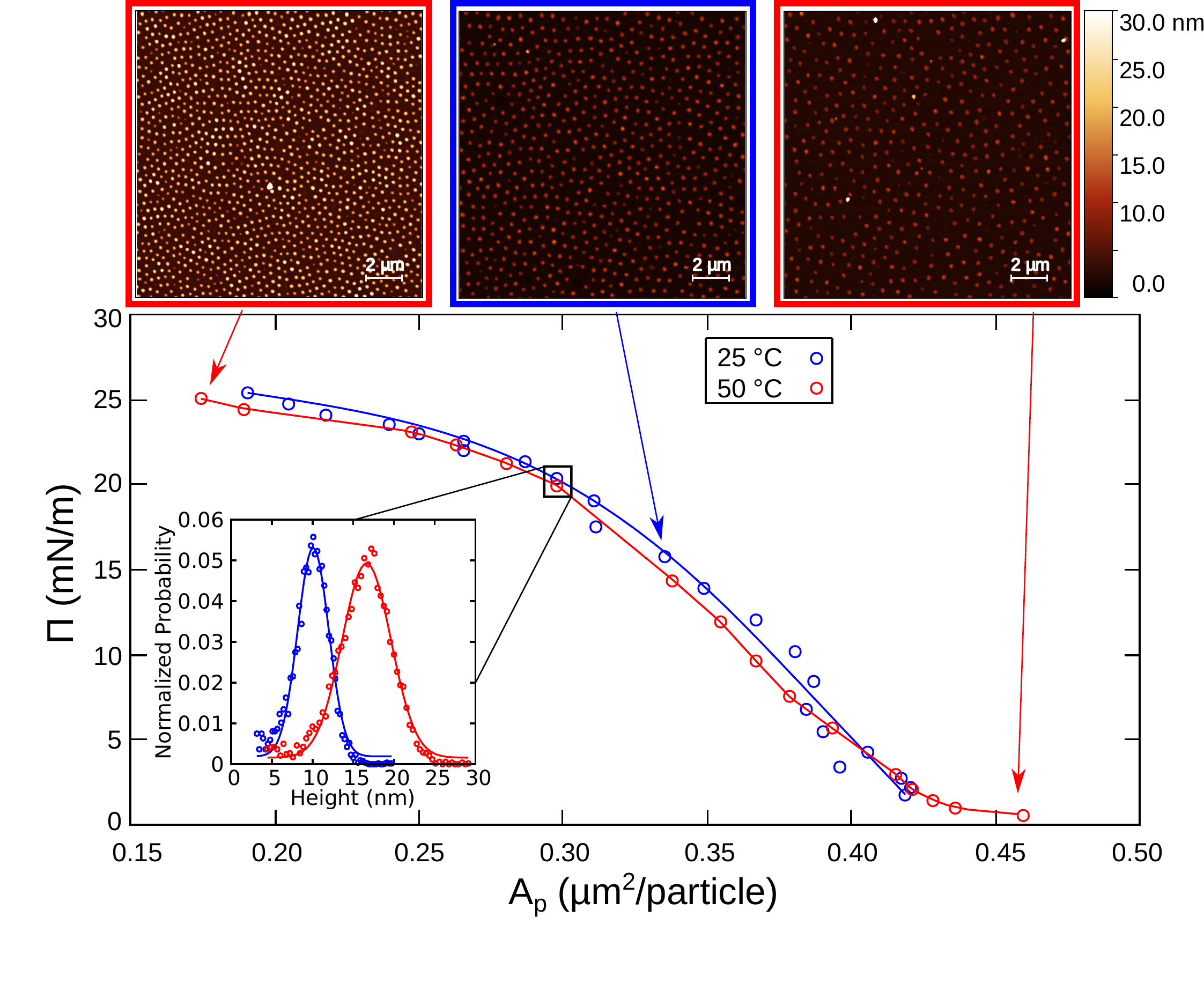}\vspace{-1.2cm}
\caption{Surface pressure $\Pi$ versus area per particle $\text{A}_{p}$ and representative AFM images obtained by the simultaneous compression and deposition of a monolayer of microgels from a water/air interface on a silicon wafer. The inset shows the microgel height distributions measured with AFM at $\Pi\simeq$20 $\text{mN m$^{-1}$}$ for 25 \degree C (\textcolor{blue}{$\bullet$}) and 50 \degree C (\textcolor{red}{$\bullet$}). The color of the images frames showcase the temperature at which microgels were deposited at. The lines are guides to the eye.}
\label{fig:Figura2}
\end{center}
\end{figure*} 

We show the influence of $\text{POM}^{3-}$ concentration on the size of microgels in Figure \ref{fig:Figura1}c. At 25 \degree C, the size of the microgels decreased from 280 nm to 240 nm at $5\cdot 10^{-5}$ M, where a minimum in $D_h$ was observed at the same $\text{POM}^{3-}$ concentration where we observed the isoelectric point in Figure \ref{fig:Figura1}b. We previously found that it is possible to deswell the pNIPAM microgels below the VPTT thanks to ionic specific interactions \cite{lopez2007hofmeister}. Nevertheless, on that study we used simpler ions belonging to Hofmeister series. Moreover, the concentration of ions at which microgels begun to deswell were always above $3\cdot10^{-1}$ M, significantly higher than those used here. By increasing the $\text{POM}^{3-}$ concentration, the adsorption of $\text{POM}^{3-}$ on the microgels resulted in their re-swelling, reaching the original $D_h$ at $5\cdot10^{-4}$ M. This happened at the corresponding $\text{POM}^{3-}$ concentration where $\mu_{e}$ showed a charge reversal in Figure \ref{fig:Figura1}b. This re-swelling was not previously observed with the ions belonging to the Hofmeister series. This points out to a stronger and more complex interaction of $\text{POM}^{3-}$ anions with the positively charged pNIPAM microgels. It is reasonable to expect that the $\text{POM}^{3-}$ anions (and $\text{Na}^{+}$ cations) absorbed within the microgel polymeric network would cause an electrostatic repulsion, resulting in their re-swelling. Furthermore, increasing the $\text{POM}^{3-}$ concentration above $10^{-3}$ M resulted again in a partial deswelling of the microgels. Our hypothesis for this non-monotonic behaviour is that this later deswelling of the microgels might come from the hydration of the $\text{POM}^{3-}$ ions, which would compete for the water molecules hydrating the pNIPAM chains, causing the microgel deswelling \cite{lopez2007hofmeister,lopez2006cationic}. 

At 50 \degree C, in the microgel deswollen state, the most relevant result was the further deswelling of the microgels at $10^{-5}$ M indicating that the microgel expelled even more water above the VPTT due to the presence of $\text{POM}^{3-}$ anions. It should be noted that this deswelling appears at a concentration at which $\mu_{e}$ was still positive in Figure \ref{fig:Figura1}b. This further $\simeq30\%$ size reduction remained constant in all the range of tested $\text{POM}^{3-}$ concentrations above $10^{-5}$ M. Previous studies with pNIPAM microgels showed a two-step deswelling when temperature was close to their VPTT \cite{lopez2006cationic, Elancheliyan2022, Giovanni2021}. In our case, the two-step deswelling occurs at 50 \degree C, well above the VPTT where the microgel is already in a clear deswollen state. To our knowledge, this is the first time that such significant $30\%$ further deswelling above the VPTT has been observed.

\

Next, we will discuss the effect of the interaction of $\text{POM}^{3-}$ anions with pNIPAM microgels confined at water/air interfaces by performing Langmuir-Blodgett experiments, where the monolayers were deposited on silicon substrates and characterized by atomic force microscopy (AFM). In these experiments, both the subphase in the Langmuir trough and the microgel dispersion were kept at pH 3 and at the same $\text{POM}^{3-}$ concentration.

Recent studies showed a plethora of new behaviors happening when microgels are confined at 2D fluid interfaces \cite{Fernandez_review}. Therefore, before analysing the role of $\text{POM}^{3-}$, we characterized the compression curve of our bare microgels at pH 3 without salt, below and above the VPTT (see Figure \ref{fig:Figura2}, and Figures \ref{fig:SupportingGrad25}-\ref{fig:SupportingGrad50}). 

The adsorption of microgels at water/air interfaces is a process out of thermodynamic equilibrium, hence our denomination of a compression curve instead of a compression isotherm. This is because slightly different amounts of deposited microgels at the interface lead to different compression curves, making comparisons difficult or meaningless when only attending to the total mass of the microgels assumed to be deposited at the interface. With our method of measuring the ex-situ area per particle for each value of $\Pi$ we avoid this problem and we can compare between different experiments and conditions.

We reproduced the behavior for small microgels that we reported in a previous work \cite{scheidegger2017compression}, where the small size was accompanied by polydispersity that frustrated the crystallization of the monolayer, as seen in the AFM images from Figure \ref{fig:Figura2}. We also reproduced the results of a previous work finding that the microgels did not change their diameter at the interface regardless of being above or below the VPTT \cite{bochenek2019effect, harrer2019stimuli}, as reflected by the overlapping compression curves at 25 and 50 \degree C. The portion immersed in the subphase was still thermoresponsive, resulting in the deswelling of that portion above the VPTT, which was reflected in an increase in height of the microgel after the deposition on the silicon substrate (see the inset in Figure \ref{fig:Figura2}). Therefore, a higher height after the deposition is a signature of deswelling of the portion immersed in the subphase before the deposition.

\begin{figure}[!ht]
\begin{center}
\includegraphics[width=\linewidth]{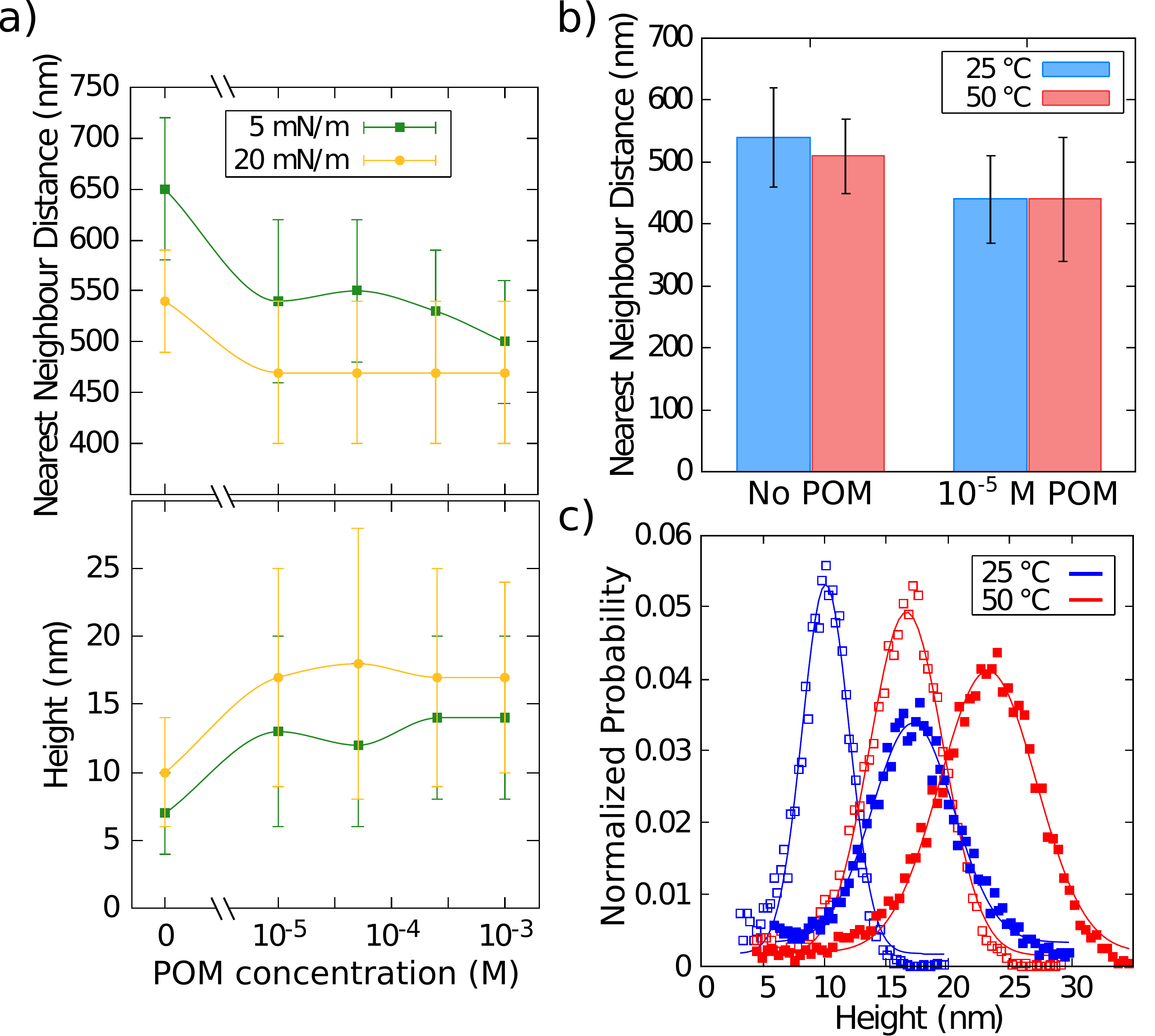}
\caption{\textbf{a)} Nearest Neighbour Distance (NND) in microgel monolayers (top) and maximum height of the microgels (bottom) vs $\text{POM}^{3-}$ concentration at $\Pi= 5$  (\textcolor{green}{$\blacksquare$}) and 20 $\text{mN m$^{-1}$}$ (\textcolor{goldenrod}{$\blacklozenge$}), at 25 \degree C. \textbf{b)} NND at $\Pi= 20$ $\text{mN m$^{-1}$}$ of microgels without $\text{POM}^{3-}$ (left) and with $10^{-5}$ M $\text{POM}^{3-}$ (right), at 25 \degree C (blue) and 50 \degree C (red). \textbf{c)} Maximum height of microgels distribution at $\Pi= 20$ $\text{mN m$^{-1}$}$, open symbols correspond to measurements in the absence of salt at 25 \degree C (\textcolor{blue}{$\square$}) and 50 \degree C (\textcolor{red}{$\square$}), while solid symbols are in the presence of $10^{-5}$ M of $\text{POM}^{3-}$ at 25 \degree C (\textcolor{blue}{$\blacksquare$}) and 50 \degree C (\textcolor{red}{$\blacksquare$}). Lines are guides to the eye.}\vspace{-0.3cm}
        \label{fig:Figura3}
\end{center}
\end{figure} 

As explained in the introduction, recent new experimental evidence has reported the role of charges on microgels confined at fluid interfaces, in this case by changing the pH of pH-responsive microgels \cite{schmidt2020influence}. Our present study aims to contribute to fill the gap on the role of ionic specificity in the behavior of microgels confined at fluid interfaces. By size, our microgels would be \say{small microgels} according to our previous study \cite{scheidegger2017compression} and the one by Schmidt et al. \cite{schmidt2020influence}.

\begin{figure*}[!t]
\begin{center}
\includegraphics[width=\linewidth]{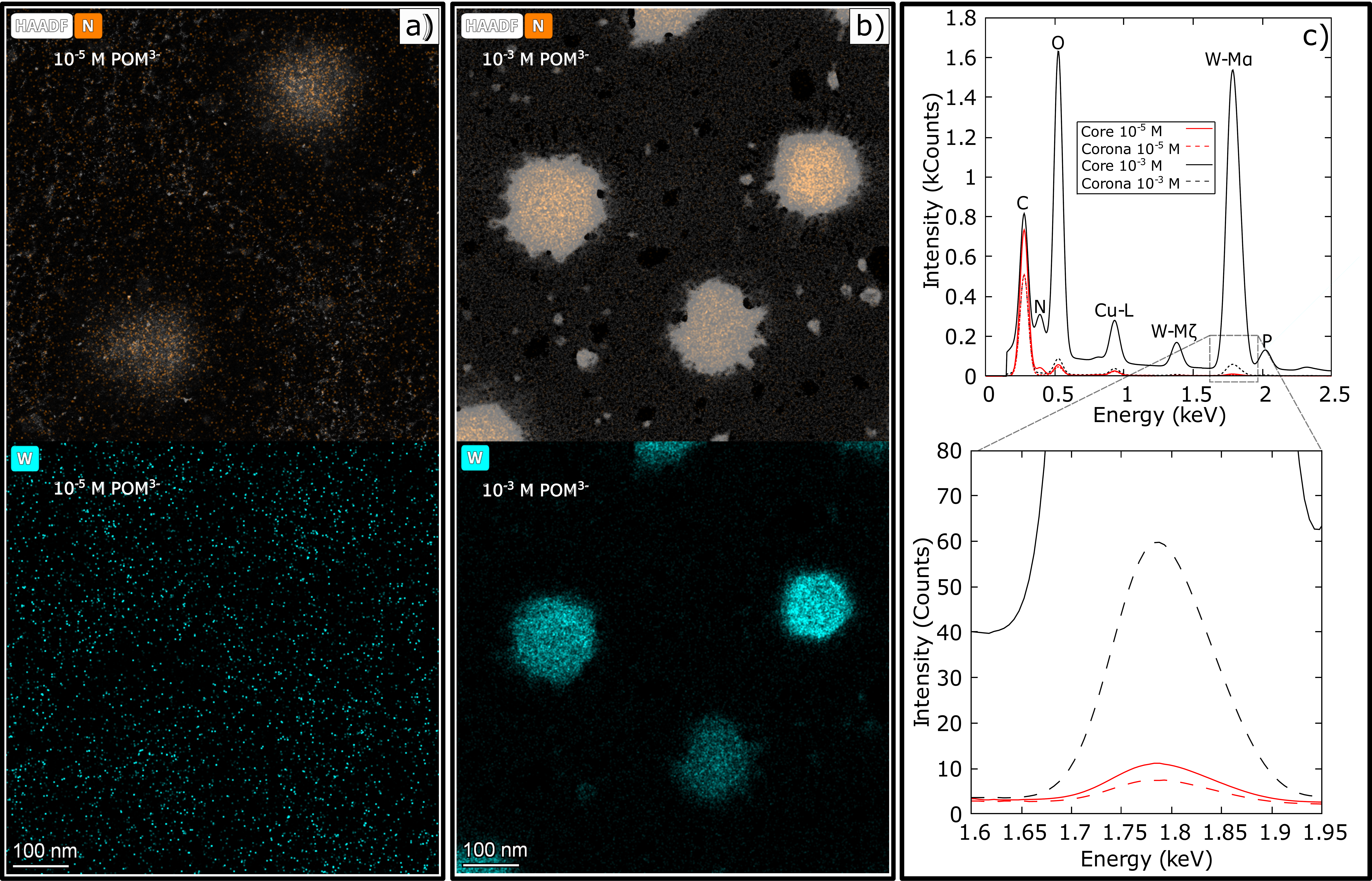}
\caption{HAADF-STEM images and EDS mapping of the microgel monolayers deposited at $\simeq$20 $\text{mN m$^{-1}$}$ with \textbf{a)} $10^{-5}$ M of $\text{POM}^{3-}$ and \textbf{b)} $10^{-3}$ M of $\text{POM}^{3-}$. Top row pictures show $N$ mapping, while bottom row pictures show $W$ mapping. \textbf{c)} Spectra obtained from EDS. Solid/dashed lines correspond to the core and corona of a deposited microgel in a-b), respectively. Red/black color corresponds to $10^{-5}$ M and $10^{-3}$ M $\text{POM}^{3-}$ concentrations, respectively. The bottom image shows a zoom-in of the region of interest for $10^{-5}$ M.}
\label{fig:Figura4}
\end{center}
\end{figure*}

In Figure \ref{fig:Figura3}a, we studied the role of the ionic specificity of the $\text{POM}^{3-}$ in the self-assembly of microgels confined at liquid interfaces at 5 and 20 $\text{mN m$^{-1}$}$ at 25 \degree C, i.e. below the VPTT. We tracked the position of each microgel in the AFM images (see SI for details), and we characterized the nearest neighbour distance (NND), which reflects the diameter of the microgels at the water/air interface. In addition, we determined the maximum height of the deposited microgel monolayers as the average of this value for all the microgels in an image. The uncertainty associated to each measurement was calculated from the Full Width at Half Maximum (FWHM) of the height and NND distributions. This uncertainty appeared as a consequence of the polydispersity of the microgels at the interface \cite{bochenek2019effect}. 

In the absence of $\text{POM}^{3-}$, in Figure \ref{fig:Figura3}a we found the expected compression of the monolayer from 5 to 20 $\text{mN m$^{-1}$}$, with the NND being reduced from 650$\pm$70 nm to 540$\pm$50 nm, and the maximum height increasing from 7$\pm$3 nm to 10$\pm$4 nm. When the $\text{POM}^{3-}$ was added at $10^{-5}$ M, the microgels at 5 $\text{mN m$^{-1}$}$ reduced their NND, from 650$\pm$70 nm to 540$\pm$50 nm, and increased their height from 7$\pm$3 to 13$\pm$7 nm. Hence, the NND and height of the deposited microgels at 5 $\text{mN m$^{-1}$}$ in the presence of $10^{-5}\, M\,\text{POM}^{3-}$ exhibit similar values as the microgels deposited at 20 $\text{mN m$^{-1}$}$ in the absence of $\text{POM}^{3-}$. In other words, the addition of $\text{POM}^{3-}$ at such low concentration as $10^{-5}$ M produced an effect on the microgel monolayer which was equivalent to mechanically compress it from 5 to 20 $\text{mN m$^{-1}$}$. As the concentration of $\text{POM}^{3-}$ increased above $10^{-5}$ M, the NND and height remained practically constant, reflecting that at $10^{-5}$ M the effect produced by the $\text{POM}^{3-}$ on the microgels at the interface already saturated. This behavior was different to the one observed for the microgels in bulk in Figures \ref{fig:Figura1}b-c, where in the range of the concentrations measured, significant variations of the $\mu_{e}$ and $D_{h}$ were observed. Interestingly, Schmidt et al. found a decrease in the diameter of small microgels at the interface upon pH-swelling \cite{schmidt2020influence}. Our results go in a different direction since the decrease in diameter in the presence of $\text{POM}^{3-}$ was accompanied by a deswelling of the portion immersed in the subphase, reflected in an increase of height after the deposition.

We also measured the effect of NaCl in the self assembly of the microgels at the interface in order to compare with the $\text{POM}^{3-}$ results. Measurements with NaCl were carried out at $10^{-1}$ M, 100 times more concentrated than the most concentrated tested $\text{POM}^{3-}$ solution. If we would like to compare ionic strengths then the NaCl concentration has to be divided by 6. We did not find any changes in either the NND nor the heights of the microgels with respect to the absence of salt (Figure \ref{fig:SupportingNaCl100}, and Table \ref{table:SupNaCl}). This evidences that the results obtained with $\text{POM}^{3-}$ come from the strong ionic specific interactions with pNIPAM microgels. This strong affinity is also reflected in an onset of effects at similar concentrations both in bulk and at the interface.

In Figure \ref{fig:Figura3}b-c, we investigated the effect of temperature for microgel monolayers in the presence of $\text{POM}^{3-}$. In the absence of $\text{POM}^{3-}$, we observed similar values of NND below and above the VPTT, as discussed in Figure \ref{fig:Figura2} \cite{schmidt2020influence, harrer2019stimuli}. Similarly, by adding $10^{-5}$ M of $\text{POM}^{3-}$, the NND remained equal below and above the VPTT, but with a lower value than in the case of no $\text{POM}^{3-}$, as already shown in Figure \ref{fig:Figura3}a. Furthermore, we observed a two-step height increase in Figure \ref{fig:Figura3}c, in accordance to the two-step deswelling shown in bulk in Figure \ref{fig:Figura1}c. First, the deswelling of the swollen core of the microgel when the temperature was above the VPTT in the absence of salt only affected the height of the deposited microgels, increasing it when compared to their corresponding depositions at 25 \degree C. Next, the addition of $\text{POM}^{3-}$ at 25 \degree C increased the height of the deposited microgels, matching the one achieved by heating the sample above the VPTT in the absence of $\text{POM}^{3-}$, as reflected by the overlapping distribution functions in Figure \ref{fig:Figura3}c. Thus, by adding $10^{-5}$ M of $\text{POM}^{3-}$ we accomplished an effect equivalent to heat above the VPTT. Furthermore, when we heated the sample in the presence of $\text{POM}^{3-}$ up to 50 \degree C, we observed a further increase in their height up to $23 \pm 9$ nm. As explained above, these results are in good agreement with the two-step deswelling observed in bulk in Figure \ref{fig:Figura1}c, where the further deswelling observed for the microgels in bulk above the VPTT reflects in a further increase of the height of the deposited microgels in Figure \ref{fig:Figura3}c.

As discussed above, $\text{POM}^{3-}$ anions are not interfacially active, but their confinement at the interface was promoted by the interfacial activity of the microgels \cite{hohenschutz2020nano,tang2009formation}. Therefore, it is reasonable to expect an accumulation of the $\text{POM}^{3-}$ anions both at the liquid interface where the coronas of the microgels are stretched, and in the respective portions still immersed in the subphase. Nevertheless, the confinement of the microgels at the liquid interface restrict their ability to re-arrange, which could cause the saturation of the effects observed. This might be compared with the constant size observed in bulk above the VPTT, where the microgel is more rigid and therefore has less ability to re-arrange. There, the $\text{POM}^{3-}$ kept adsorbing on the microgel when the concentration was increased, as reflected in the decrease of $\mu_{e}$ between $10^{-5}$ M and $10^{-4}$ M in Figure \ref{fig:Figura1}b. However, this increased adsorption of the $\text{POM}^{3-}$ was not reflected on the size in Figure \ref{fig:Figura1}c, where the deswelling effect produced by the $\text{POM}^{3-}$ already saturated at $10^{-5}$ M.

In order to locate the distribution and amount of $\text{POM}^{3-}$ in the pNIPAM microgels, we characterized the location of the pNIPAM and $\text{POM}^{3-}$ anions on the deposited monolayers by HAADF-STEM and EDS (see details in Materials, SI and Figures \ref{fig:SupportingPreparationTEM}, \ref{fig:SupportingTEM5500x}, and \ref{fig:SupportingTEM}). Since the $\text{POM}^{3-}$, $[PW_{12}O_{40}]^{3-}$, contains wolframium ($W$), we used it as an indicator of the $\text{POM}^{3-}$ concentration, while nitrogen ($N$) was used as an indicator of the presence of pNIPAM. It is important to note that the STEM sample preparation resulted in the imaging of a single monolayer of microgels. In Figure \ref{fig:Figura4}a-b, we present the HAADF-STEM and EDS mapping of the microgels deposited at $\Pi\simeq$20 $\text{mN m$^{-1}$}$ in the presence of $10^{-5}$ and $10^{-3}$ M of $\text{POM}^{3-}$, respectively. In both Figures the denser and more cross-linked core revealed a higher $N$ concentration than in the surrounding stretched corona. While it is difficult to see by eye in Figure \ref{fig:Figura4}a, Figure \ref{fig:Figura4}b shows that the $\text{POM}^{3-}$ was accumulated both at the microgel core and the stretched corona, with more $\text{POM}^{3-}$ on the core compared to the stretched corona. To illustrate this preferential adsorption on the core of the microgel, Figure \ref{fig:Figura4}c shows the corresponding EDS spectra in the core of one of the microgels compared to its corona (see more details in Figures \ref{fig:SupportingTEM001All}, \ref{fig:SupportingTEM005All}, and \ref{fig:SupportingTEM1All}). In the zoom-in, it is possible to see how the signal corresponding to the $W-M\alpha$ peak of the EDS spectra is greater in the core region of the microgel also for $10^{-5}$ M of $\text{POM}^{3-}$.

These results confirm that at $10^{-3}$ M there was more $\text{POM}^{3-}$ accumulation compared to $10^{-5}$ M. Nevertheless, the effects produced on the deswelling of the portion of the microgel immersed in the subphase, and on the decrease of the diameter at the interface, already saturated at $10^{-5}$ M, as shown in Figure \ref{fig:Figura3}a through the NND and height of the deposited monolayers characterized by AFM. Hence, the sensitivity of these techniques enabled us to distinguish the accumulation of $\text{POM}^{3-}$ in the microgel monolayers as its concentration increased.

\begin{table}[t!]
\begin{center}
	\caption{XPS analysis with atomic percentages of the elements present in a substrate with microgels deposited at $\Pi = 20$ $\text{mN m$^{-1}$}$ at 25 \textdegree C with $10^{-5}$ M $\text{POM}^{3-}$. The relative concentrations of each peak from the deconvoluted spectra of $W4f$ is also shown, where $W^{(5+)}$ corresponds to partially degraded $\text{POM}^{3-}$.}
 \label{table:XPS}
\vspace{0.3cm} 
\begin{tabular}{cc}
\hline
\begin{tabular}[c]{@{}c@{}}Signal\\ (Binding Energy, eV)\end{tabular} & (\%)                                                             \\ \hline
$C1s$ (285.03)                                                                 & 28.25                                                                     \\
$N1s$ (399.92                                                                  & 3.7                                                                       \\
$O1s$ (532.21)                                                                 & 33.1                                                                      \\
$Si2p$ (98.99)                                                                 & 33.9                                                                      \\
$W4f$                                                                          & 0.7                                                                       \\ \hline
\begin{tabular}[c]{@{}c@{}}XPS fitting\\ of the W4f Spectra\end{tabular}       & \begin{tabular}[c]{@{}c@{}}(\% relative to\\ the W4f signal)\end{tabular} \\ \hline
$W^{(6+)} 4f_{7/2}$ (35.95)                                                    & 41.0                                                                      \\
$W^{(6+)} 4f_{5/2}$ (38.00)                                                    & 30.8                                                                      \\
$W^{(6+)} 4p_{3/2}$ (40.91)                                                    & 5.9                                                                       \\
$W^{(5+)} 4f_{7/2}$ (34.45)                                                    & 12.7                                                                      \\
$W^{(5+)} 4f_{5/2}$ (36.44)                                                    & 9.6                                                                       \\ \hline
\end{tabular}
\end{center}
\end{table}

Moreover, we also observed some results that point out a restructuring of the microgel polymer network by the presence of $\text{POM}^{3-}$, an effect observed for free pNIPAM chains in bulk \cite{buchecker2019self}. For example, in Figure \ref{fig:SupportingTEM}, the bridging of two microgel cores is accompanied by the presence of a higher concentration of $\text{POM}^{3-}$ in the pNIPAM bridge. This bridging was never observed in the absence of $\text{POM}^{3-}$ (see Figures \ref{fig:Figura2}, \ref{fig:SupportingGrad25} and \ref{fig:SupportingGrad50}). Therefore, compared to free pNIPAM chains in bulk \cite{buchecker2019self}, the restructuring ability of the $\text{POM}^{3-}$ would be significantly reduced when interacting with microgels due to the cross-linked polymer network, but it seems to be still present (see Figure \ref{fig:SupportingEstructuracion}).

\

We further studied the integrity of the $\text{POM}^{3-}$ anions after the deposition of a microgel monolayer from the water/air interface onto the silicon substrates by XPS (see details in the SI). The presence of $W^{(6+)}4f_{7/2}$ and $W^{(6+)}4f_{5/2}$ doublet peaks at binding energies of 35.95 and 38.00 eV revealed the presence of unaltered $WO_{3}$, indicating that the $\text{POM}^{3-}$ structure remained unaltered \cite{cheng2019self}. A slight reduction into the $W^{5+}$ oxidation state was observed as new $W^{(5+)}4f_{7/2}$ and $W^{(5+)}4f_{5/2}$ signals appeared at 34.45 and 26.44 eV, respectively, signaling the partial degradation of the $\text{POM}^{3-}$ (see Figure \ref{fig:SupportingXPSSpectrum}). 
We present the relative concentration of each element for the analysed sample in Table \ref{table:XPS} at $10^{-5}$ M of $\text{POM}^{3-}$, showing that $\simeq22\%$ of the $\text{POM}^{3-}$ partially degraded during the harsh process of depositing and drying on a silicon substrate.

\

\begin{figure}[ht!]
\begin{center}
\includegraphics[width=\linewidth]{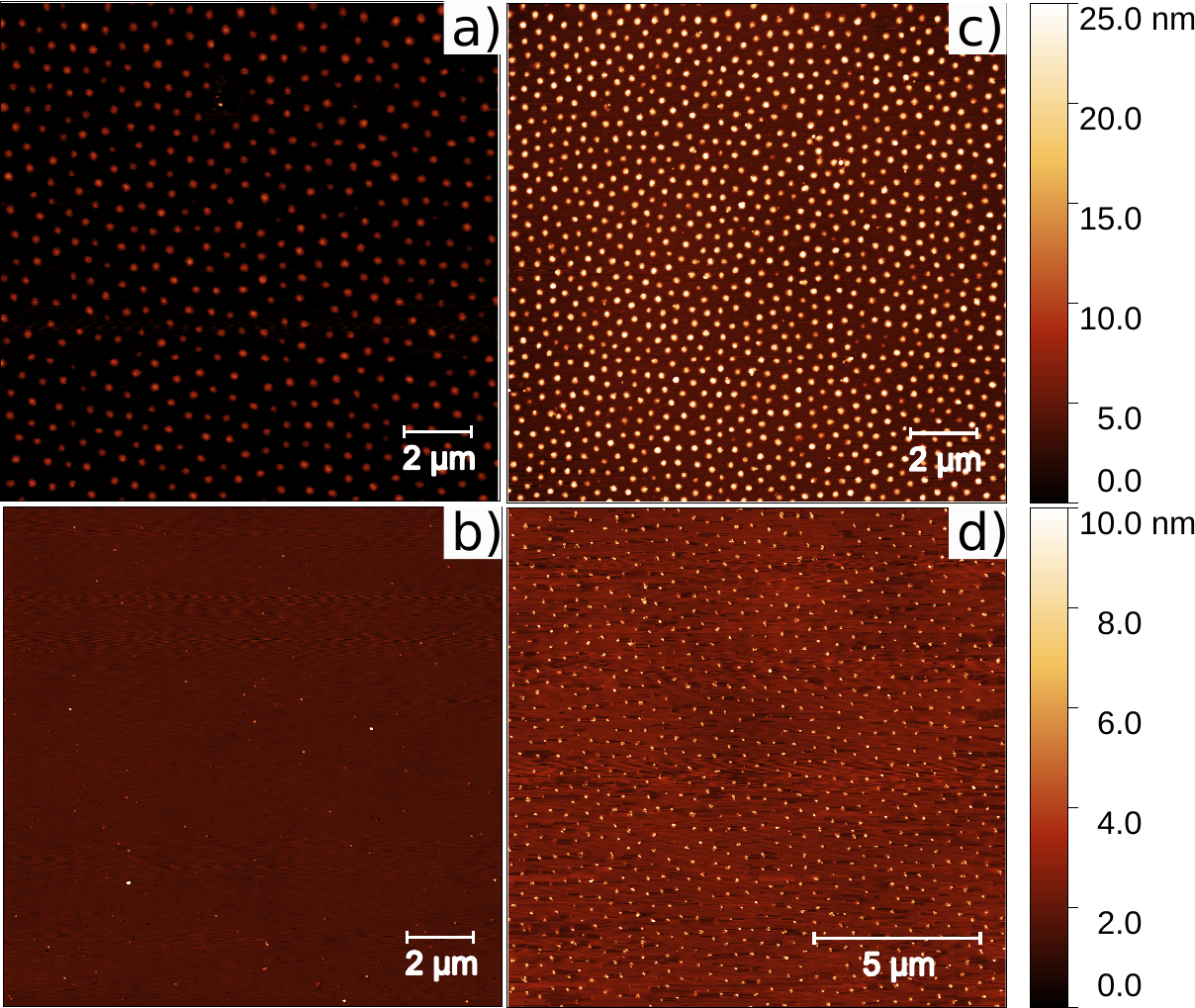}
\caption{Showcase of the improved ability of microgels to perform as lithography masks by resisting more to an air plasma treatment of 30 min at 100 W. AFM images obtained by the deposition of microgels at 20 $\text{mN m$^{-1}$}$. Depositions were done in the absence (a-b) and presence (c-d) of $\text{POM}^{3-}$ at $10^{-5}$ M. AFM images were acquired before (a-c) and after (b-d) the plasma treatment.}
        \label{fig:Figura5}
\end{center}
\end{figure} 

Finally, to showcase if the microgels restructured by $\text{POM}^{3-}$ improve their ability to perform as lithography masks, we subjected microgel monolayers to plasma etching to compare their resistance to be etched. To do this, substrates with microgels deposited at $\Pi= 20$ $\text{mN m$^{-1}$}$ with no $\text{POM}^{3-}$ and with $10^{-5}$ M $\text{POM}^{3-}$ were subjected to air plasma at 100 W for 30 min. We acquired AFM images before and after the treatment. In Figure \ref{fig:Figura5}, it is visible the compression of the monolayer induced by the $\text{POM}^{3-}$, with a reduction of the NND and brighter, i.e. higher, microgels. Height distribution profiles of the micrographs in Figure \ref{fig:Figura5} are shown in Figure \ref{fig:AlturaPlasma}. After the plasma treatment, we noticed that the addition of $10^{-5}$ M $\text{POM}^{3-}$ prevented the full etching of the microgel monolayer. This would hint to an improved performance as lithography masks by plasma dry etching. During this process, it is the complete etching of the microgels acting as lithography masks what determines the upper boundary for the height of the produced vertically aligned nanowires \cite{fernandez2021near}. Thus, more resistance of the lithography masks to plasma etching would result in higher vertically aligned nanowires during a dry etching process.
         
\section*{Conclusion}
\label{conclusion}

In this paper we show evidences of the strong interaction between positively charged pNIPAM microgels and Keggin-type $[\text{PW}_{12}\text{O}_{40}]^{3-}$ anions ($\text{POM}^{3-}$), both in bulk and at the water/air interface. 

Ionic specificity effects were observed at low concentrations of $\text{POM}^{3-}$ (ca. $5\cdot10^{-5}$ M), and below and above the VPTT of the microgels. In bulk, we observed charge inversion in all ranges of temperatures investigated, specially in the deswollen state. The hydrodynamic diameter $D_h$ was also affected; Below the VPTT, it exhibited a deswelling-swelling-deswelling trend with the $\text{POM}^{3-}$ concentration. Above the VPTT, we observed a novel \textit{extra} $30\%$  deswelling of the microgel at $10^{-5}$ M of $\text{POM}^{3-}$. These novel results are a consequence of the great interaction of $\text{POM}^{3-}$ with the pNIPAM microgels, as it has been reported for free pNIPAM chains in bulk \cite{buchecker2019self}. We also conclude the important role of the softness of the microgels below and above the VPTT, with effects that cannot be explained considering the microgels above the VPTT as hard particles \cite{lyon2012polymer}.

At the water/air interface, the interaction of $10^{-5}$ M $\text{POM}^{3-}$ with the microgel monolayers below the VPTT resulted in a decrease of the diameter of the microgels at the interface and a deswelling of the portion immersed in the subphase. The later was reflected in a height increase of the deposited microgels. In the absence of $\text{POM}^{3-}$, it was necessary to compress from 5 to 20 $\text{mN m$^{-1}$}$ to achieve an equal reduction in the diameter, and to heat above the VPTT to observe similar values in the increased height.
Above the VPTT, we only observed an extra deswelling of the portion immersed in the subphase when $\text{POM}^{3-}$ was added, in agreement with the results in bulk. The monolayer behavior did not change above $10^{-5}$ M of $\text{POM}^{3-}$, indicating a saturation effect, despite the further accumulation of $\text{POM}^{3-}$ on the pNIPAM characterized by EDS mapping. We propose an explanation for this saturation effect and the one observed in bulk for the $D_h$ above the VPTT. In both cases, the microgel ability to re-arrange, i.e. its softness, is restricted either due to the confinement at the water/air interface or due to the deswelling above the VPTT in bulk. This restriction in the softness would be responsible for the observed saturation effects. 

As a general conclusion, this work points out the importance of the ionic specific interactions over the tuning of the bulk behaviour and 2D interfacial self-assembly of pNIPAM microgels below and above the VPTT. It stands out that the behaviour of soft matter in presence of ionic specificity effects is richer and more complex than those with hard colloidal particles \cite{faraudo2023interaction}. Hence, to obtain a deeper understanding at molecular level on the interaction between $\text{POM}^{3-}$ and pNIPAM microgels, further experiments including molecular dynamics simulations are required. In addition, we have already envisioned a possible application of this type of interaction in the improvement of soft colloidal lithography, since the microgel monolayers showed improved resistance to plasma etching. 

\section*{Acknowledgements}
We acknowledge Dr. Jordi Faraudo Gener and Dr. Carlos Drummond for insightful conversations, the CIC from University of Granada for the STEM-HAADF and EDS measurements, and the SCAI from the University of Málaga for the XPS measurements. This work was supported by the projects PID2020-116615RA-I00 funded by MCIN/AEI/ 10.13039/501100011033, and EMERGIA grant with reference EMC21\_00008, and projects PY20-00241, A-FQM-90-UGR20 funded by Consejer\'ia de Universidad, Investigaci\'on e Innovaci\'on de la Junta de Andaluc\'ia, and by FEDER “ERDF A way of making Europe”.

\section*{Author contributions}
\textbf{Antonio Rubio-Andrés}: Investigation, Data curation, Validation, Visualization, Writing- Original draft preparation. \textbf{Delfi Bastos-González}: Conceptualization, Methodology, Resources, Supervision, Writing- Original draft preparation. \textbf{Miguel Angel Fernandez-Rodriguez}: Conceptualization, Methodology, Software, Resources, Supervision, Writing- Original draft preparation, Project administration, Funding acquisition.

\setlength{\bibsep}{0.0cm}
\bibliography{example_refs}

\clearpage

\onecolumngrid
\section*{Graphical Abstract}
\begin{center}
    \includegraphics[width=\linewidth]{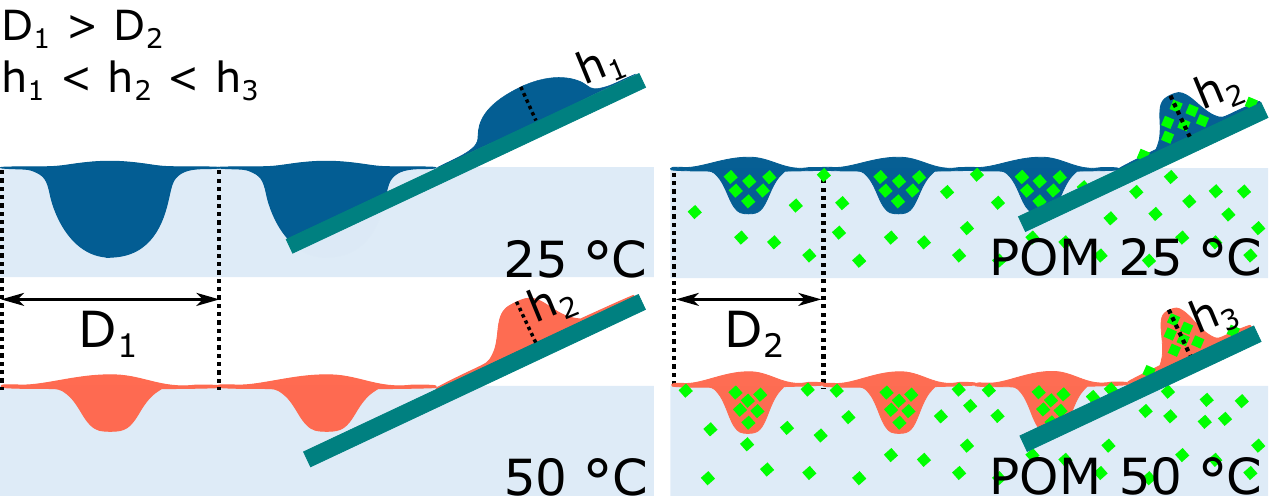} 
\end{center}
\section*{Highlights}
\begin{itemize}

\item Low polyoxometalate (POM) concentrations ($5\cdot10^{-5}\text{M}$) induce charge inversion in pNIPAM microgels below and above the Volume Phase Transition Temperature (VPTT). 
\item Microgels exhibit a novel two-step deswelling: one due to the VPTT and a further one above the VPTT induced by the POM.
\item Adding POM to a microgel monolayer at a liquid interface is similar to compress and heat it above the VPTT due to restructuration.
\item There is a saturation effect of the POM added to microgels confined at liquid interfaces at $10^{-5}\text{M}$. 
\item Microgels with POM exhibit an improved resistance to plasma etching, with potential applications for soft colloidal lithography.
\end{itemize}

\newpage

\renewcommand{\thefigure}{\textbf{S\arabic{figure}}}
\setcounter{figure}{0}    
\onecolumngrid

\begin{center}
    \LARGE{\textbf{Supplementary Information}}
\end{center}

 \section*{Physicochemical characterization in bulk}  

    \
    
\begin{figure}[ht!]
        \begin{center}

        \includegraphics[width=0.9\linewidth]{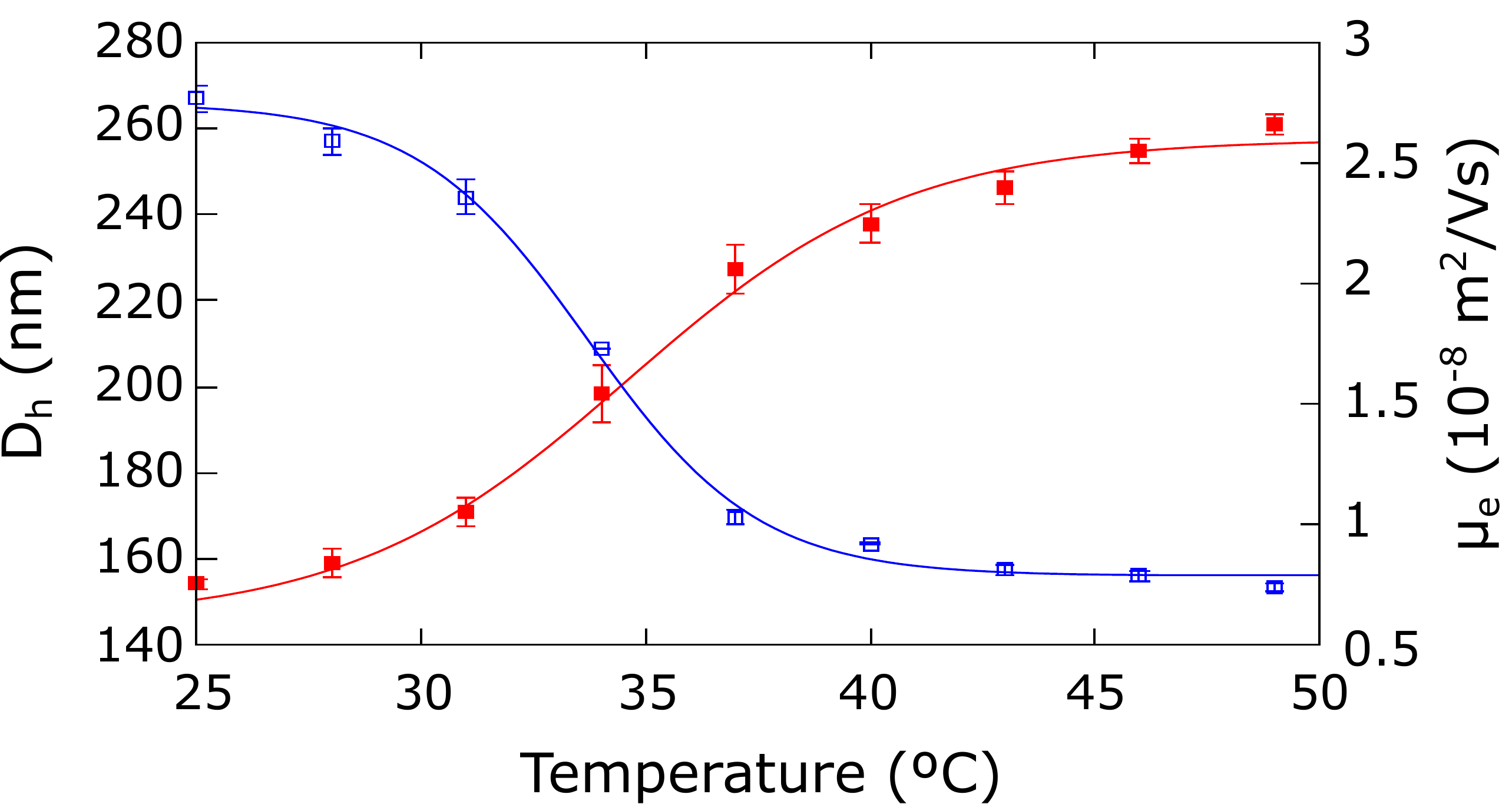}
        \caption{ Bulk characterization of the positively charged microgels. Hydrodynamic diameter $D_{h}$ (\textcolor{blue}{$\square$}) and electrophoretic mobility $\mu_{e}$ (\textcolor{red}{$\blacksquare$}) as a function of the temperature. Blue line corresponds to the sigmoidal fit of Equation \ref{eq:S1}, while red line is a guide to the eye.}
        \label{fig:SupportingCaracterizacion}
        \end{center}
    \end{figure}

    In order to further characterize the thermoresponsive behavior of the microgels, their $D_{h}$ can be modeled by the following sigmoidal equation \cite{navarro2011modified}:
    \begin{equation}\tag{S1}\label{eq:S1}
         D_h(T)=D_{collapsed}+\frac{D_{swollen}-D_{collapsed}}{1+\exp{\frac{T-\rm{VPPT}}{\tau}}}
     \end{equation}

    where $D_{swollen}$ and $D_{collapsed}$ are the $D_h$ values under and above the VPTT, respectively, and $\tau$ is the decay time of the sigmoid. Fitting this equation to the $D_{h}$ data of Figure \ref{fig:SupportingCaracterizacion} gives a VPTT for our microgels of $33.7 \pm 0.2$ \degree C.
    
    


    \newpage
    
    \section*{Langmuir trough depositions}

    \ 
    
    The microgel dispersion was diluted to 0.1 wt$\%$ with the corresponding $NaCl$ or $\text{POM}^{3-}$ solution at pH3. Isopropyl alcohol was added at 4:1 water/isopropanol ratio. The Langmuir trough has a Teflon block provided with a trough, with a maximum area of 238.5 cm$^{2}$ compressible to 58 cm$^{2}$ via two motorized Delrin barriers.  
    
    The trough was filled with the desired electrolyte solution at pH 3 while maintaining the barriers in open position and the Wilhelmy plate in contact with the interface. After 10 minutes, the Wilhelmy plate was completely wetted. Then, the interface was cleaned by closing the barriers and aspirating the water surface with a pipette tip. Then, the barriers were opened and $\Pi$ was re-zeroed. The compression + aspiration cycles were repeated until $\Pi \leq 0.3$ $\text{mN m$^{-1}$}$. A transparent Plexiglas case covered the trough to avoid any dust deposition or air flux which could disturb the measurements.


    In all experiments the barriers were compressed at a constant rate of $5.2 \: \text{mm min$^{-1}$}$. Next, the dipper arm was raised at a constant velocity of $0.5 \: \text{mm min$^{-1}$}$, either during or after the barrier compression, depending on the experiment. Once the substrate raised over the interface, it was left to air dry before being imaged in the AFM. After finishing each deposition, the trough and barriers were cleaned by rinsing with tap water, distilled water, gently wiping with Kimtech paper and isopropanol, distilled water, and MilliQ water.

\newpage
    \section*{Characterization of deposited monolayers: AFM, particle tracking, XPS, TEM, HAADF-STEM, EDS, and air plasma measurements.}

    \subsection*{AFM and particle tracking}

 \begin{figure}[ht!]
        \centering
        \includegraphics[width=\linewidth]{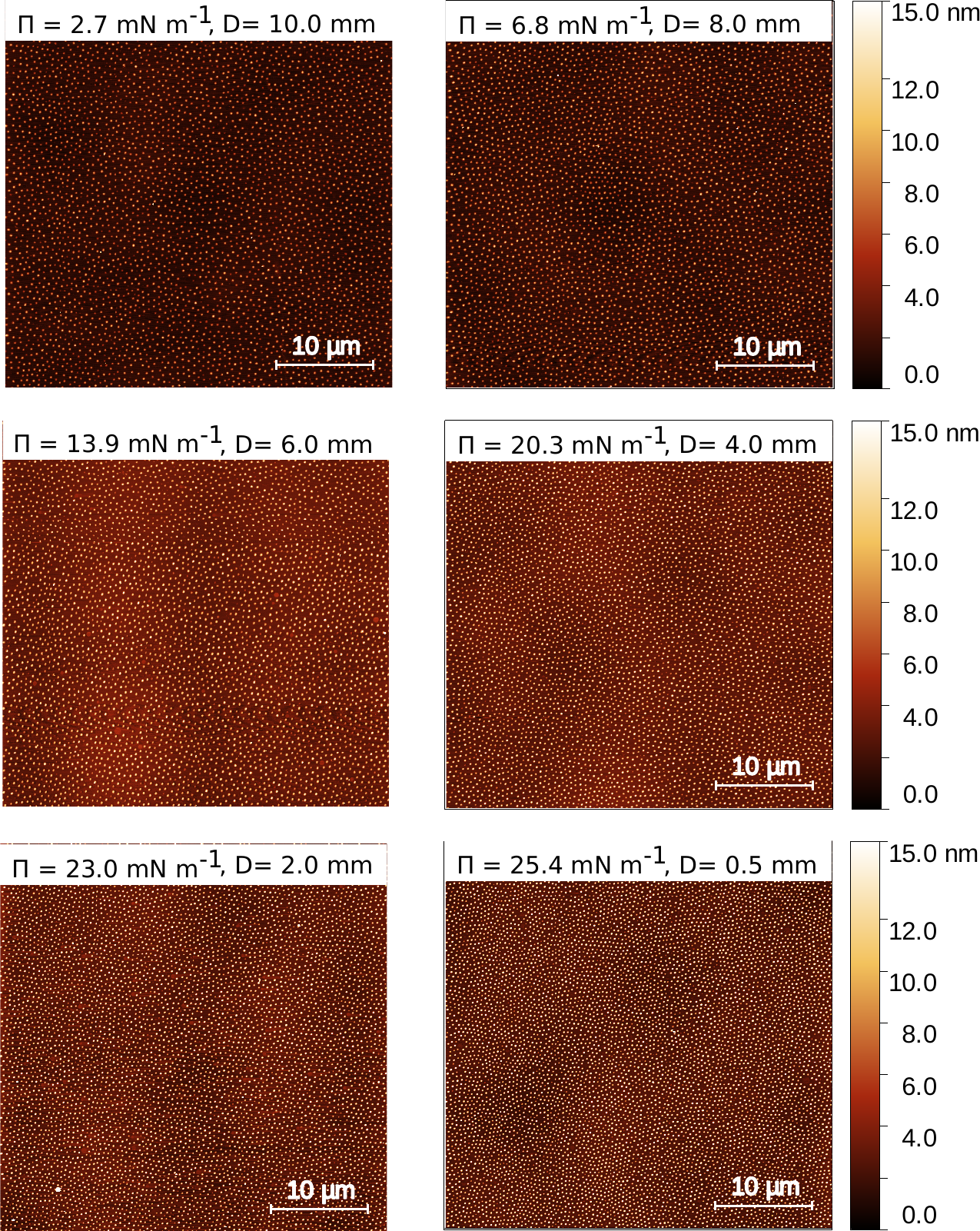}
        \caption{AFM images of the deposited microgels during the gradient deposition at 25 \degree C, used to reconstruct the compression curve of Figure \ref{fig:Figura2}. For each AFM image we indicate the corresponding $\Pi$ and distance $D$ from the reference point where we open the barriers, at the end of the deposition.}
        \label{fig:SupportingGrad25}
    \end{figure}

    \newpage  
    
     \begin{figure}[ht!]
        \centering
        \includegraphics[width=\linewidth]{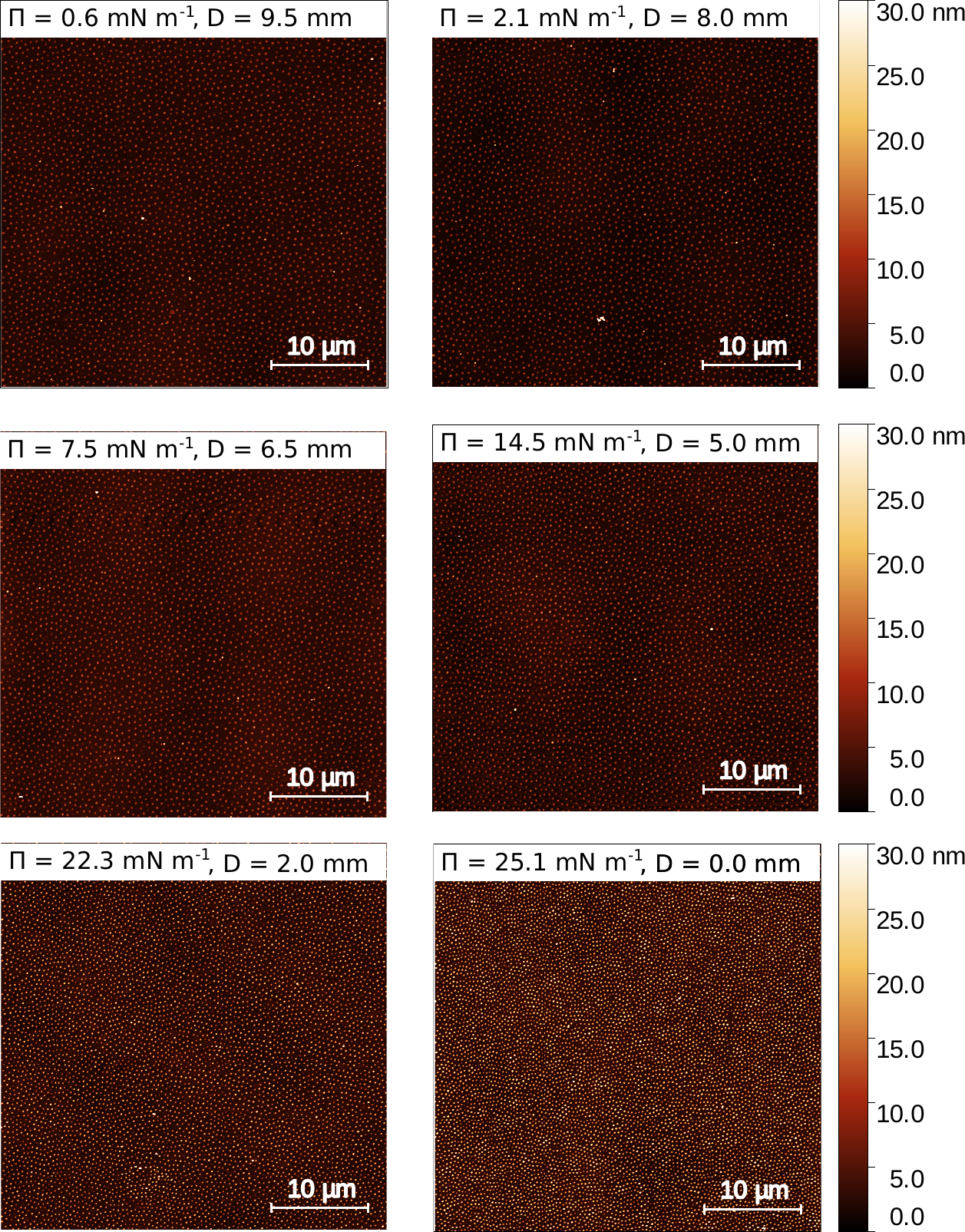}
        \caption{AFM images of the deposited microgels during the gradient deposition at 50 \degree C, used to reconstruct the compression curve of Figure \ref{fig:Figura2}. For each AFM image we indicate the corresponding $\Pi$ and distance $D$ from the reference point where we open the barriers, at the end of the deposition.}
        \label{fig:SupportingGrad50}
    \end{figure}
\clearpage
\subsection*{NaCl Measurements}

    \begin{figure}[ht!]
        \centering
        \includegraphics[width=\linewidth]{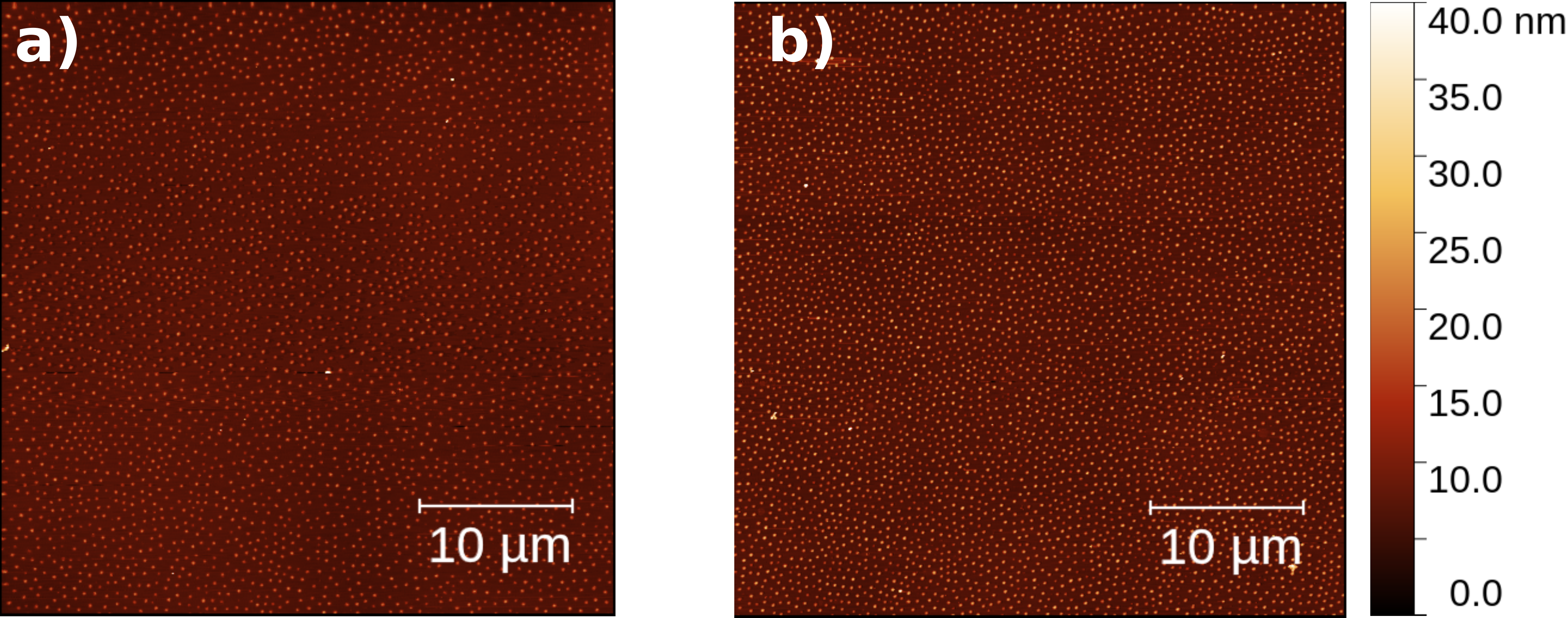}
        \caption{AFM images of the deposited microgels at \textbf{a)} 5 $\text{mN m}^{-1}$ and \textbf{b)} 20 $\text{mN m}^{-1}$ with $0.1 \: M$ of NaCl.}
        \label{fig:SupportingNaCl100}
    \end{figure}

\begin{table*}[ht!]
	\begin{center}
	\caption{Nearest Neighbour Distance (NND) between microgels and maximum height of the microgels in the absence of salt and with $0.1 \: M$ of NaCl.}\label{table:SupNaCl}
		\begin{tabular}{lcc}
\hline
\multicolumn{1}{c}{$\Pi = 5 \: \text{mN m}^{-1}$} & \text{\begin{tabular}[c]{@{}c@{}}Nearest Neighbour Distance (nm)\\ \end{tabular}} & \text{\begin{tabular}[c]{@{}c@{}}Height (nm)\end{tabular}}\\ 
\hline
\text{No salt} & 650 $\pm$ 70 & 7 $\pm$ 3\\
\text{$0.1 \: M \: $ NaCl} & 660 $\pm$ 80 & 8 $\pm$ 4\\\hline
\multicolumn{3}{l}{$\Pi = 20 \: \text{mN m}^{-1}$}\\\hline
\text{No salt} & 540 $\pm$ 50 & 10 $\pm$ 4\\
\text{$0.1 \: M \: $ NaCl} & 570 $\pm$ 70 & 11 $\pm$ 5\\\hline
\end{tabular}
\end{center}

\end{table*}

\clearpage
\subsection*{TEM characterization and EDS chemical identification}

    \begin{figure}[ht!]
        \centering
        \includegraphics[width=0.8\linewidth]{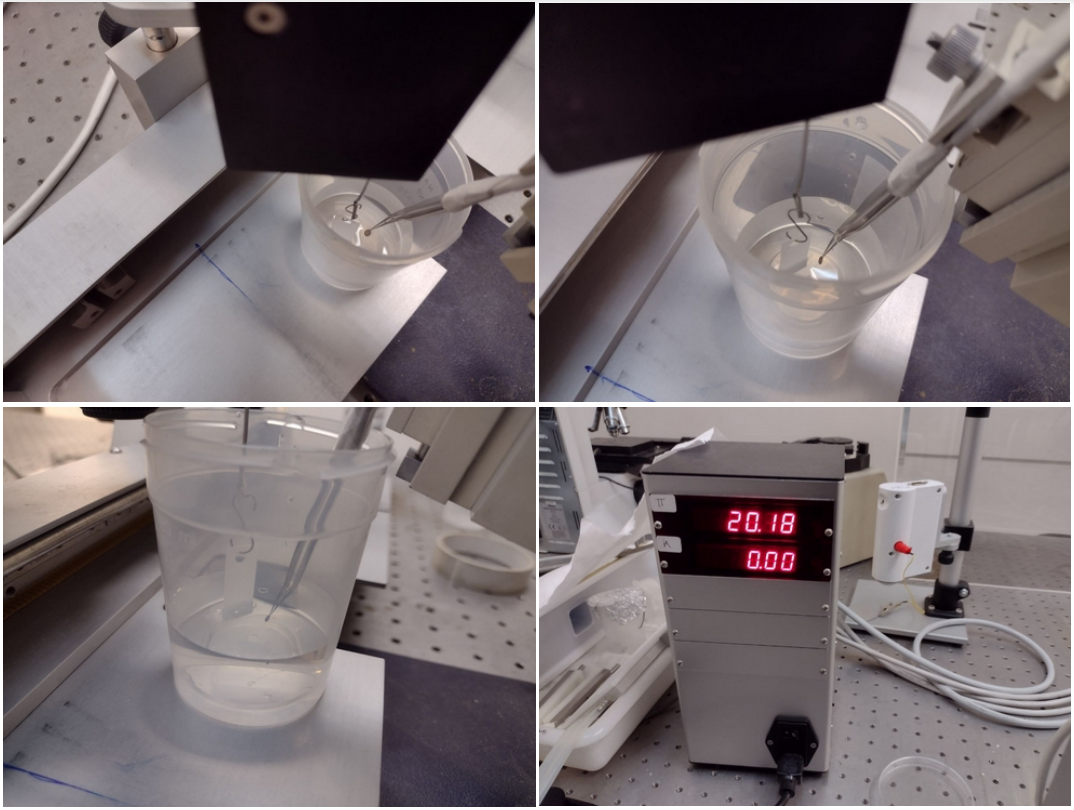}\vspace{-0.15cm}
        \caption{TEM grid preparation. We prepared a plastic container as shown in the pictures with the corresponding $\text{POM}^{3-}$ solution at pH 3, monitoring $\Pi$. Next, the TEM grid was attached to the dipper arm and placed below the interface. To ensure that the interface was clean before the microgel deposition, the interface was aspirated several times with a pipette tip connected to a vacuum pump. After this, the microgel dispersion was deposited on the interface with a microsyringe until the surface pressure reached the value $\Pi \simeq 20$ $\text{mN m}^{-1}$. Finally, the TEM grid was lifted with a constant velocity of 0.5 $\text{mm min}^{-1}$.}
        \label{fig:SupportingPreparationTEM}
        \end{figure}

        \begin{figure}[ht!]
        \centering
        \includegraphics[width=\linewidth]{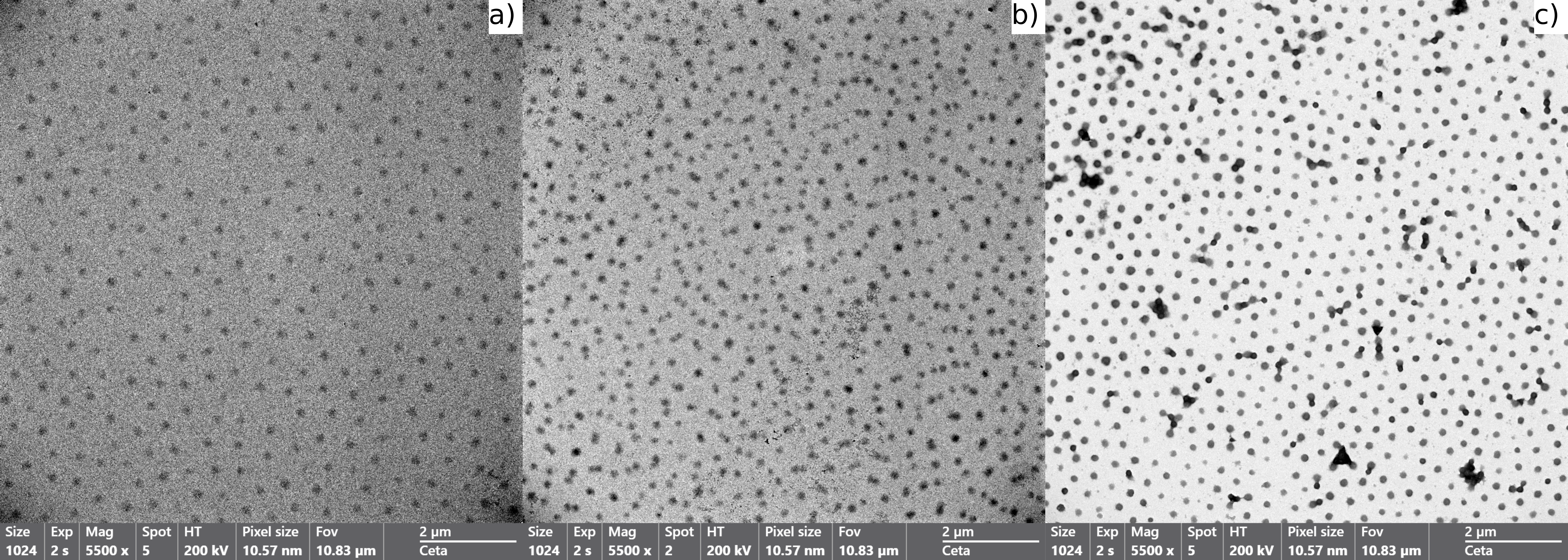}
        \caption{TEM images of the microgels deposited at $\Pi\simeq20\,\text{mN m}^{-1}$ with $\text{POM}^{3-}$ at \textbf{a)} $10^{-5}$ M, \textbf{b)} $5\cdot10^{-5}$ M and \textbf{c)} $10^{-3}$ M.}
        \label{fig:SupportingTEM5500x}
    \end{figure}
\newpage
        \begin{figure}[ht!]
        \centering
        \includegraphics[width=\linewidth]{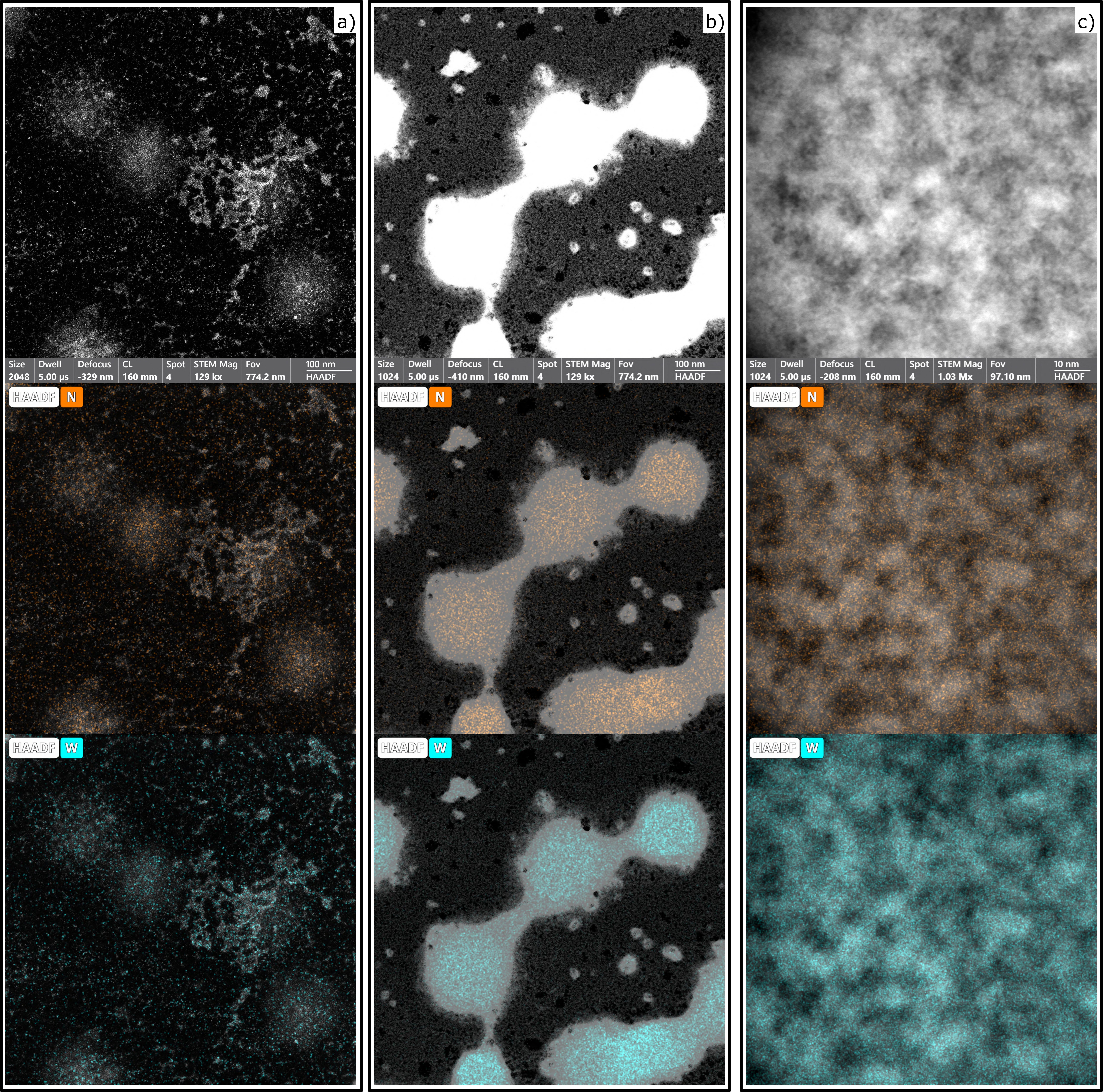}
        \caption{HAADF-STEM, and $N$ and $W$ EDS mapping of the microgel monolayers deposited at 20 $\text{mN m$^{-1}$}$ with different $\text{POM}^{3-}$ concentrations. Image of the microgels, and corresponding EDS mapping deposited at \textbf{a)} $5\cdot10^{-5}$ M, and \textbf{b-c)} $10^{-3}$ M of $\text{POM}^{3-}$, respectively. The image in a) shows clusters of $W$ above the microgels. The image in b) shows the restructuring ability of the $\text{POM}^{3-}$ that form bridges of pNIPAM and $W$, not observed in the absence of $\text{POM}^{3-}$ in Figures \ref{fig:SupportingGrad25} and \ref{fig:SupportingGrad50}. The image in c) is a zoom in the core of a microgel in b) showing the clustering of pNIPAM and $\text{POM}^{3-}$.}
        \label{fig:SupportingTEM}
    \end{figure}
\newpage

    \begin{figure}[ht!]
        \centering
        \includegraphics[width=\linewidth]{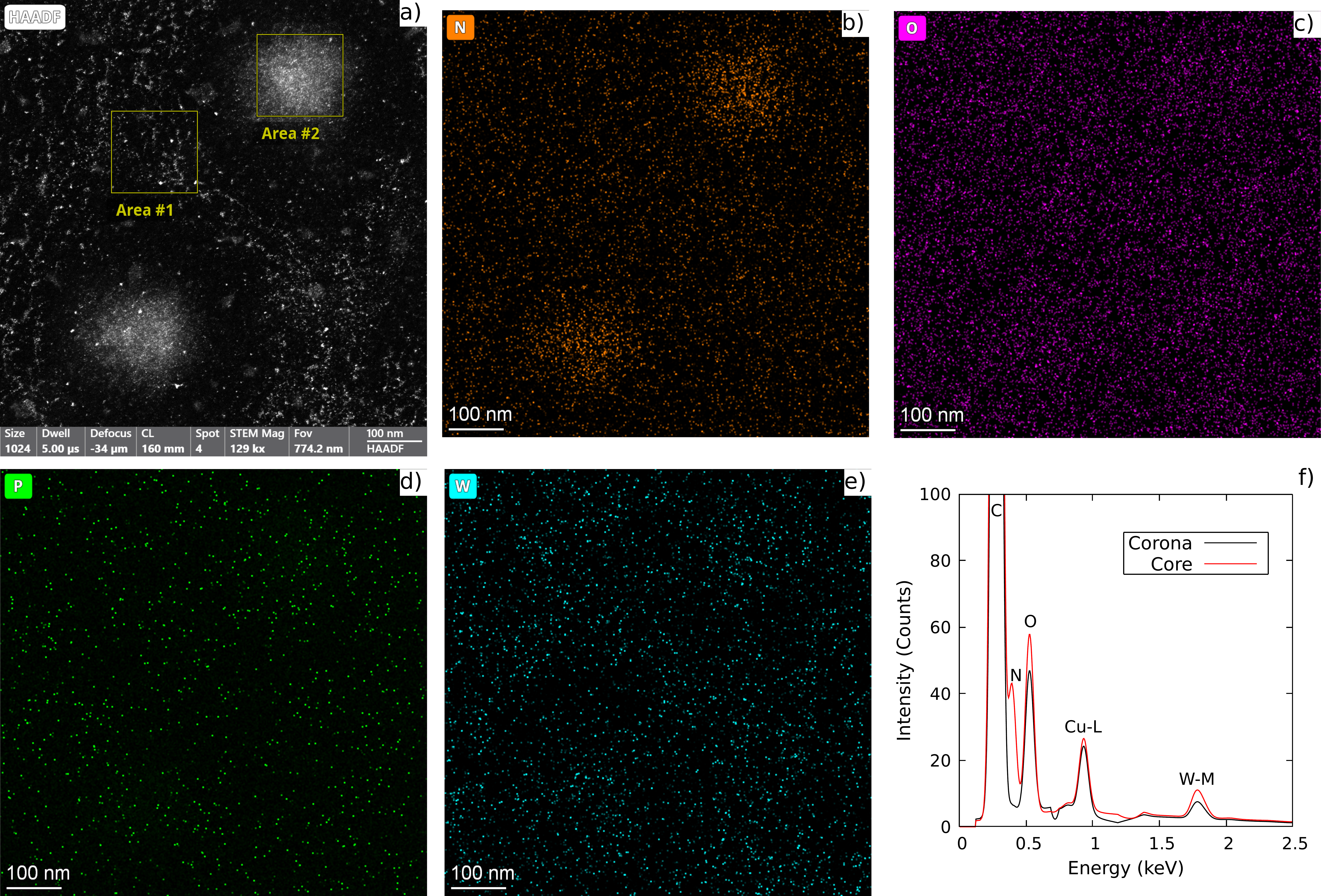}
        \caption{HAADF-STEM images and EDS mapping of the microgel monolayers deposited at $\simeq$20 $\text{mN m$^{-1}$}$ with $10^{-5}\, M$ of $\text{POM}^{3-}$. \textbf{a)} HAADF-STEM image of two microgels where the region corresponding with the corona is shown as \textit{Area \#1} and the region corresponding with the core is shown as \textit{Area \#2}. \textbf{b)-e)} EDS mapping of $N$, $O$, $P$ and $W$. \textbf{f)} Spectrum distribution obtained from EDS. Black lines correspond to the corona, while red lines correspond to the core, i.e. the \textit{Areas \#1} and \textit{\#2} in a), respectively. Although it is not easy to see in e) by eye, the EDS spectra shows more $W$ in the core than in the corona.}
        \label{fig:SupportingTEM001All}
    \end{figure}
\newpage

\begin{figure}[ht!]
        \centering
        \includegraphics[width=\linewidth]{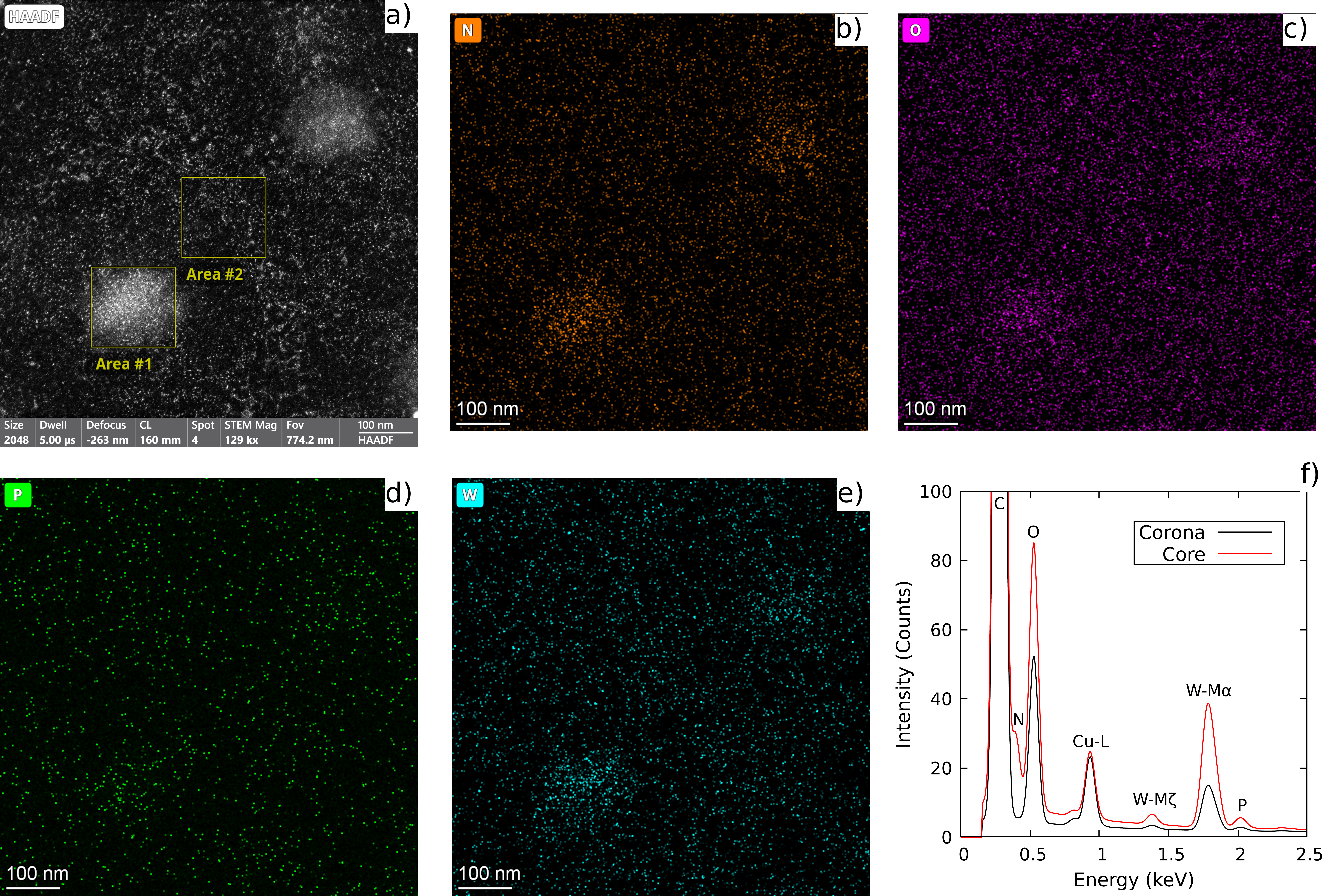}
        \caption{Same as Figure S8 but deposited with $5\cdot10^{-5}\, M$ of $\text{POM}^{3-}$.}
        \label{fig:SupportingTEM005All}
    \end{figure}
\newpage
    \begin{figure}[ht!]
        \centering
        \includegraphics[width=\linewidth]{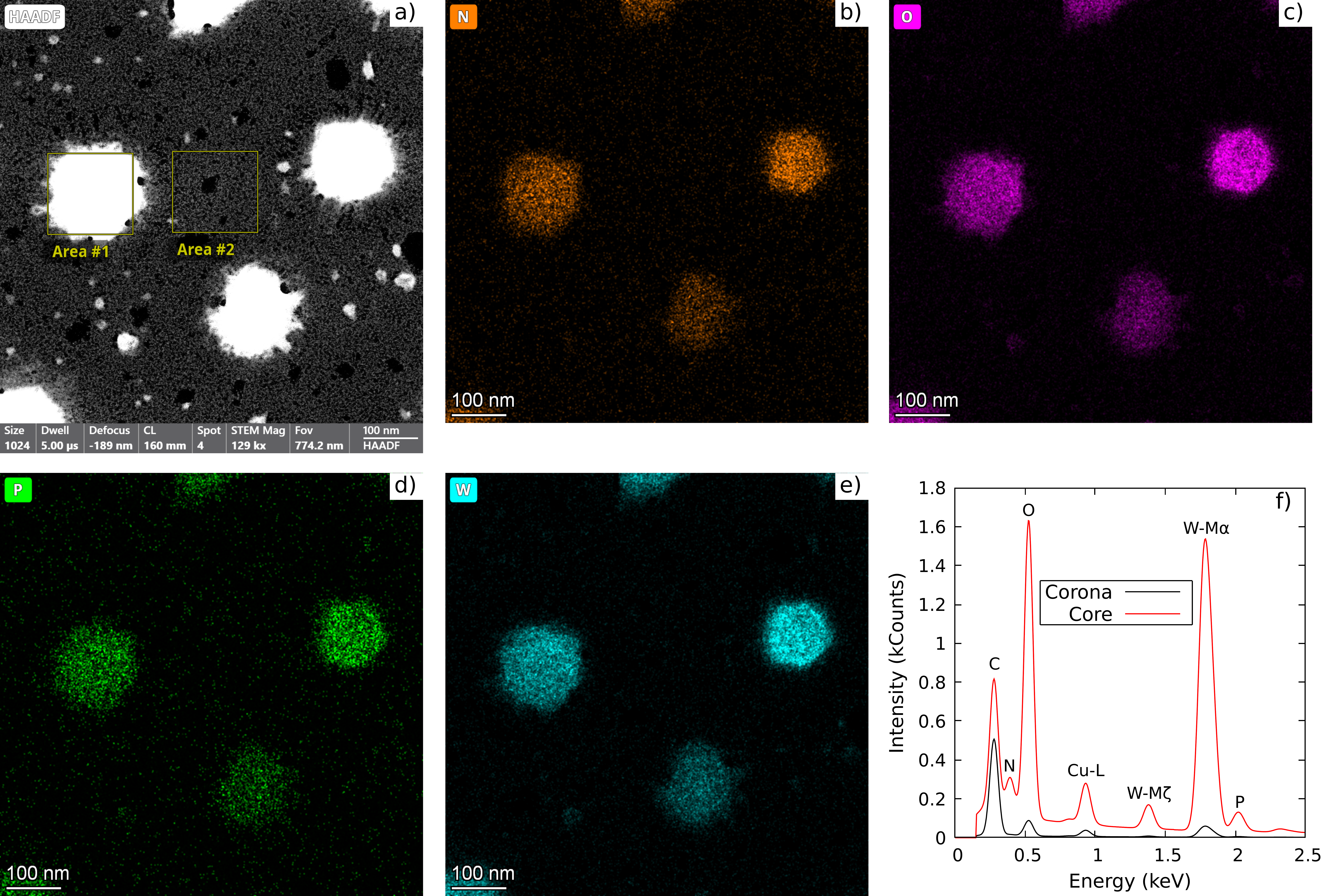}
        \caption{figure}{Same as Figure S8 and S9 but deposited with $10^{-3}\, M$ of $\text{POM}^{3-}$. The intensity of the $W-M\alpha$ peak is 12 times greater than the $P$ peak in the core at 1 mM $\text{POM}^{3-}$ concentration. Given that the $[PW_{12}O_{40}]^{3-}$ contains 12 atoms of $W$ for each $P$ atom, this result supports that entire $\text{POM}^{3-}$ anions are being adsorbed at the core. }
        \label{fig:SupportingTEM1All}
    \end{figure}

\newpage 

\begin{figure}[ht!]
        \centering
        \includegraphics[width=\linewidth]{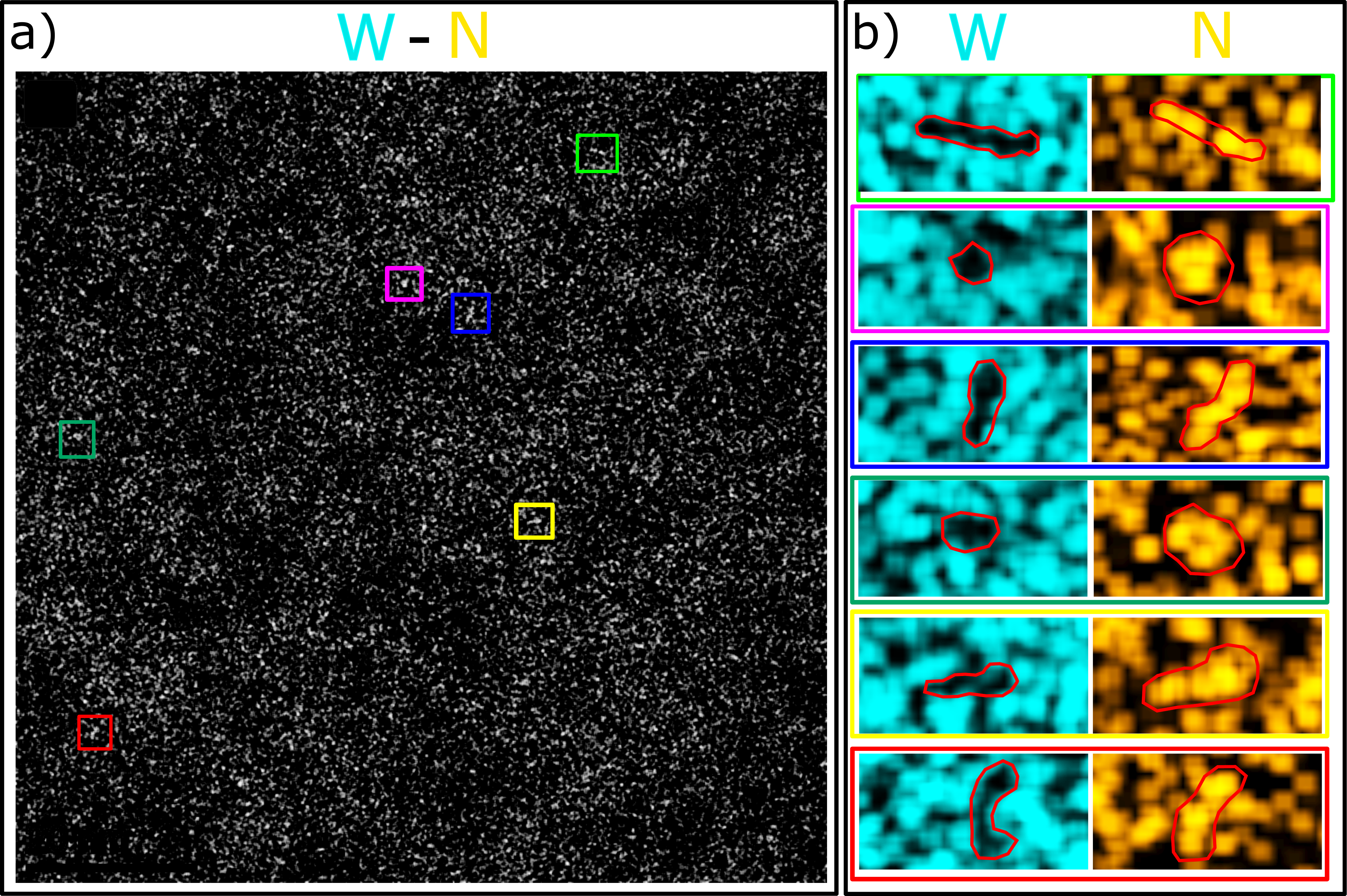}
        \caption{HAADF-STEM images and EDS mapping showing the re-structuring of $\text{POM}^{3-}$ anions around pNIPAM chains for the microgels corresponding to Figure \ref{fig:SupportingTEM}c. \textbf{a)} Subtraction of $W$ and $N$ EDS mapping images to identify possible re-structured regions. \textbf{b)} Zoom-in of the regions of interest indicated on a) with colored rectangles, with red highlighted areas as guides to the eye. Left and right columns correspond to $W$ and $N$ EDS mapping, respectively. The $W$ signal seems to be re-structured around the $N$ signal in those highlighted regions.}
        \label{fig:SupportingEstructuracion}
    \end{figure}

In Figure \ref{fig:SupportingEstructuracion} we present a HAADF-STEM + EDS mapping corresponding to the Figure \ref{fig:SupportingTEM}c, the measurement of the core of a microgel with $10^{-3}\, M$ of $\text{POM}^{3-}$. Each $W$ dot has an approximate size of 1 nm, which might point out to a single $\text{POM}^{3-}$ anion \cite{buchecker2019self}, while each $N$ dot refers to a collection of NIPAM monomers within the polymer network. By observing each region of interest shown in Figure \ref{fig:SupportingEstructuracion}, it is possible to see how $W$ dots (i.e. $\text{POM}^{3-}$ anions) are placed around $N$ dots (i.e. pNIPAM) forming structures that might resemble the ones found for free pNIPAM chains structured in the presence of $\text{POM}^{3-}$ \cite{buchecker2019self}. In our case the cross-linked pNIPAM chains are not free to re-arrange to the same extent as the free pNIPAM chains. 

\newpage

\subsection*{XPS measurements}
\begin{figure}[ht!]
        \centering
        \includegraphics[width=\linewidth]{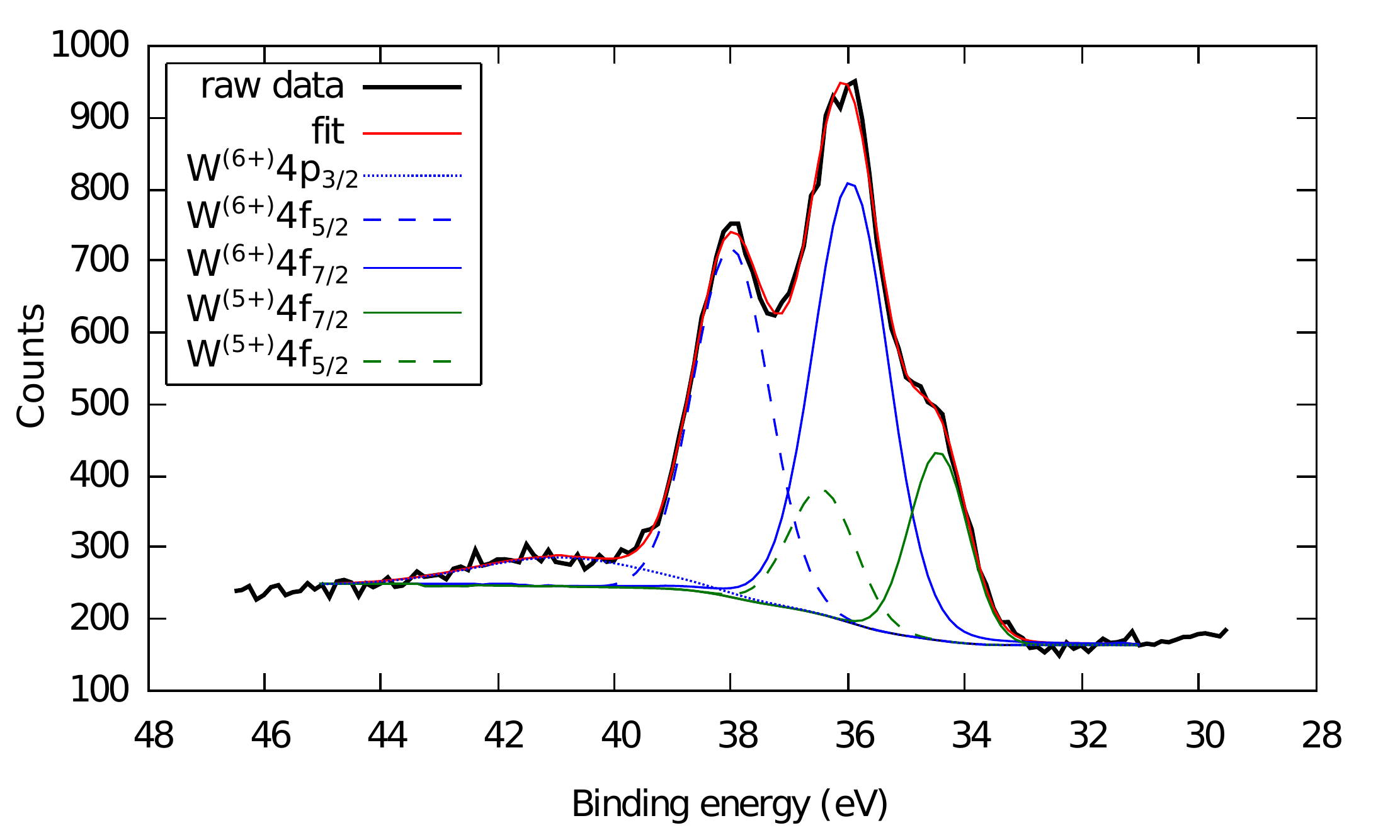}
        \caption{Deconvoluted $W4f$ spectra corresponding to the substrate with microgels deposited in the presence of $10^{-5}\, M$ of $Na_{3}PW_{12}O_{40}$.  }
        \label{fig:SupportingXPSSpectrum}
    \end{figure}

       The objective of this measurements was to check whether the $\text{POM}^{3-}$ maintained its integrity during the deposition process on the substrate. This was done by checking the $W4f$ signal and deconvoluting it in its different peaks, as presented in Figure \ref{fig:SupportingXPSSpectrum}. The $W4p_{3/2}$ peak (blue dotted line), together with the $W4f_{5/2}$ and $W4f_{7/2}$ doublet peaks (blue dashed and solid lines) placed at binding energies of 38.00 and 35.95 eV, respectively, correspond to $W^{6+}$ in the expected $\text{POM}^{3-}$ anion state \cite{gautam2022three}. The doublet corresponding to $W^{5+}$ (green lines), shows that W has been reduced and the anion degraded changing its chemical composition.

\newpage

\subsection*{Plasma etching}

\begin{figure}[ht!]
        \centering
        \includegraphics[width=\linewidth]{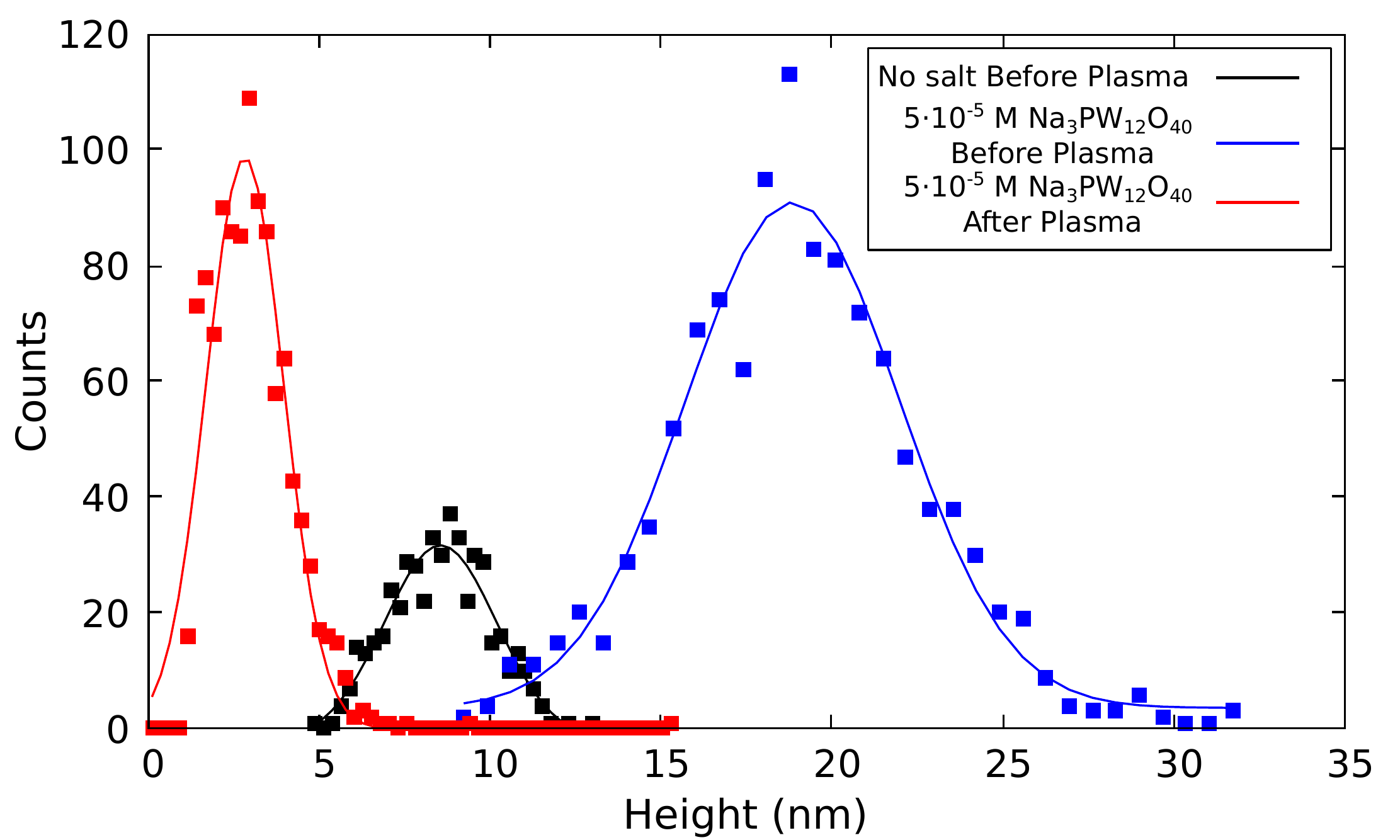}
        \caption{Height distribution corresponding to the microgels deposited in the absence of salt (black) and with $5\cdot10^{-5} \: M$ of $Na_{3}PW_{12}O_{40}$ before (blue) and after (red) being subjected to air plasma ashing for 30 min at 100 W. The microgels with no salt completely disappeared after the plasma treatment.}
        \label{fig:AlturaPlasma}
    \end{figure}

\end{document}